\pdfoutput=1
\documentclass[journal=jctcce,manuscript=article]{achemso}
\usepackage{graphicx}
\usepackage{hyperref}
\usepackage{subfigure}
\usepackage{algorithmic}
\usepackage{algorithm}
\usepackage{appendix}
\usepackage{rotating}
\usepackage{array}
\usepackage{bm}
\usepackage{dcolumn}
\usepackage{amsmath}
\usepackage{txfonts}
\usepackage{color}
\usepackage{physics}
\usepackage{braket}
\usepackage{xfrac}
\DeclareMathOperator*{\argmin}{argmin}

\usepackage{xcolor}
\hypersetup{
    colorlinks,
    linkcolor={red!50!black},
    citecolor={blue!50!black},
    urlcolor={blue!80!black}
}
\usepackage{tikz}
\usetikzlibrary{shapes.geometric, arrows}

\newcommand{\Emat}[1]{\ensuremath{\{\mathbf{E}^{(#1)}\}}}
\newcommand{\Evec}[1]{\ensuremath{\mathbf{E}^{(#1)}}}
\newcommand{\Fmat}[1]{\ensuremath{\{\mathbf{F}^{(#1)}\}}}

\newcommand{\EmatX}[1]{\ensuremath{\{\mathbf{E}^{(#1)}\mathbf{x}\}}}

\newcommand{\vecX}{\ensuremath{\mathbf{x}}}
\newcommand{\crop}[1]{\ensuremath{\hat{c}^{\dagger}_{#1}}}
\newcommand{\anop}[1]{\ensuremath{\hat{c}_{#1}}}

\tikzset{font={\fontsize{9pt}{10.5pt}\selectfont}}
\tikzstyle{startstop} = [rectangle, rounded corners, minimum width=1cm, minimum height=0.5cm, text centered, draw=black, fill=red!10]
\tikzstyle{io} = [trapezium, trapezium left angle=70, trapezium right angle=110, minimum width=1cm, minimum height=0.5cm, text centered, text width=3cm, draw=black, fill=orange!10]
\tikzstyle{process} = [rectangle, minimum width=1cm, minimum height=0.5cm, text centered, draw=black, fill=blue!10, text width=2cm, execute at begin node=]
\tikzstyle{vprocess} = [rectangle, minimum width=1cm, minimum height=0.5cm, text centered, draw=red, fill=blue!10, text width=2cm, line width=2pt, execute at begin node=]
\tikzstyle{decision} = [diamond, text centered, draw=black, aspect=1, fill=green!10, minimum height=0.5cm, execute at begin node=]
\tikzstyle{vdecision} = [diamond, text centered, draw=red, aspect=1, fill=green!10, minimum height=0.5cm, line width=2pt, execute at begin node=]
\tikzstyle{arrow} = [thick,->,>=stealth]

\title{The Variational Localized Active Space Self-Consistent Field Method}
\date{\today}
\author{Matthew R.\ Hermes}
\email{herme068@umn.edu}
\author{Laura Gagliardi}
\email{gagliard@umn.edu}
\affiliation{Department of Chemistry, Chemical Theory Center, and The Minnesota Supercomputing Institute, University of Minnesota, Minneapolis, MN, 55455}

\begin{document}

\begin{abstract}
Fragmentation methods applied to multireference wave functions constitute a road towards the application of highly accurate \emph{ab initio} wave function calculations to large molecules and solids. However, it is important for reproducibility and transferability that a fragmentation scheme be well-defined with minimal dependence on initial orbital guesses or user-designed \emph{ad hoc} fragmentation schemes. One way to improve this sort of robustness is to ensure the energy obeys a variational principle; i.e., that the active orbitals and active space wave functions minimize the electronic energy in a certain \emph{ansatz} for the molecular wave function. We extended the theory of the localized active space self-consistent field, LASSCF, method (\textit{JCTC} \textbf{2019}, \textit{15}, 972) to fully minimize the energy with respect to all orbital rotations, rendering it truly variational. The new method, called vLASSCF, substantially improves the robustness and reproducibility of the LAS wave function compared to LASSCF. We analyze the storage and operation cost scaling of vLASSCF compared to orbital optimization using a standard CASSCF approach and we show results of vLASSCF calculations on some simple test systems. We show that vLASSCF is energetically equivalent to CASSCF in the limit of one active subspace, and that vLASSCF significantly improves upon the reliability of LASSCF energy differences, allowing for more meaningful and subtle analysis of potential energy curves of dissociating molecules. We also show that all forms of LASSCF have a lower operation cost scaling than the orbital-optimization part of CASSCF.
\end{abstract}
\maketitle

\section{Introduction \label{sec:intro}}

Current targets of quantum chemical simulation such as lanthanide\hskip0pt/\hskip0ptactinide\hskip0pt-\hskip0ptligand complexes\cite{Hallmen2018,Hallmen2019,Sharma2019} and metal-organic frameworks,\cite{Horike2009,Lee2009a,Odoh2015,Coudert2016,Yan2017,Bernales2018} frequently contain a large number of strongly-correlated electronic degrees of freedom in their wave functions, which renders traditional multiconfiguration self-consistent field (MC-SCF) approaches like the complete active space self-consistent field (CASSCF) method\cite{Roos1980} problematic due to the factorial explosion of the computational cost of the latter with respect to the size of the active space. However, in many of these systems, the strongly-correlated electronic degrees of freedom are centered around separable units, for example distant transition metal nuclei\cite{Pandharkar2019} or weakly-entangled monomers,\cite{Nishio2019} and a low-scaling local correlation\cite{Wang2019b} or fragmentation\cite{Gordon2012} approach to the MC-SCF framework may generate realistic chemical models.

Along these lines, we recently introduced the localized active space self-consistent field (LASSCF) method\cite{Hermes2019} for strongly-correlated systems characterized by weakly-entangled subunits. LASSCF was originally conceived as a generalization of density matrix embedding theory (DMET),\cite{Knizia2012,Wouters2016,Wouters2017} but unlike the latter, LASSCF generates a true wave function for the whole molecule and provides an upper bound to CASSCF and full configuration interaction (FCI) energies. LASSCF produces a localized active space (LAS) wave function, which is an approximation to a CASSCF wave function in which the active space is split into one or more non-overlapping, unentangled subspaces. This approximation to CASSCF eliminates the inherent factorial operation and storage cost explosion with increasing system size that is associated with handling a single configuration interaction (CI) vector spanning a direct-product basis of orbitals. Initial tests of LASSCF showed that it reproduces the results of comparable CASSCF calculations, as long as the strong electron correlation of the test system was localized and short-range; whereas the more general DMET method using a CASSCF solver fails dramatically.\cite{Hermes2019} Further tests showed that LASSCF is an attractive alternative to CASSCF in determining spin-state energy ladders for organometallic compounds, especially multinuclear compounds.\cite{Pandharkar2019}

LASSCF as explored in Refs.\ \citenum{Hermes2019} and \citenum{Pandharkar2019} is not truly variational in the sense that not all of its wave function parameters minimize the energy. One consequence of this is that analytical molecular gradients are less straightforward to implement. Although analytical gradients are usually still possible, by using, for instance, Lagrange's method of undetermined multipliers,\cite{Press1992} this requires that the method be represented as a constrained energy minimization. In the case of LASSCF, the constraints on the orbital optimization are provided indirectly by the initial guesses for the active orbitals as well as the orbital-localization scheme used to split the atomic orbital (AO) space of the molecule into non-overlapping fragments. Strong dependence of total energies and other observables on these initialization parameters is observed even for systems with only one physically reasonable state. Not only does this make the implementation of analytical molecular gradients more problematic, it impairs the reproducibility of total energies and density matrices calculated with the LASSCF method. Initial guess states are manifold, especially in the field of MC-SCF calculation, where workers are accustomed to exploring a wide variety of \emph{ad hoc} protocols for generating initial guesses in the pursuit of elusive states;\cite{Schmidt1998,Veryazov2011,Sayfutyarova2017,Bao2019,Khedkar2019} a useful MC-SCF method should depend on the initial guess only inasmuch as the choice of the guess should allow the user to select a desired state from among a finite number of energy minima in the wave function parameter space.

Here we present an extension to LASSCF which we call ``variational LASSCF'' or vLASSCF in order to distinguish it from the method described in Ref.\ \citenum{Hermes2019}, and which fully minimizes the energy with respect to all possible transformations of the orbitals and CI vectors of a LAS wave function. This extension improves the robustness and reproducibility of the LAS wave function and energy, makes the method energetically equivalent to CASSCF and Hartree--Fock (HF) in the limits respectively of one active subspace and one determinant in the active superspace, and does not incur substantial additional computational cost. Going forward, vLASSCF should replace LASSCF entirely.

Equivalents to the LAS \emph{ansatz} have been explored before in the literature. For example, the LAS wave function can be understood as a form of the cluster mean-field wave function explored by Jim\'{e}nez-Hoyos and Scuseria in the context of 1D or 2D Hubbard model systems,\cite{Jimenez-Hoyos2015} or as a ``rank-one basis state'' defined by Nishio and Kurashige in their studies of molecular aggregates,\cite{Nishio2019} or as the bond dimension $=1$ case of an active-space decomposition (ASD) wave function\cite{Parker2013,Parker2014,Parker2014a,Kim2015} obtained using the density matrix renormalization group (DMRG) extension.\cite{Parker2014b} Among these works, the cluster mean-field method described in Ref.\ \citenum{Jimenez-Hoyos2015} and the \emph{ad hoc} protocol employed for one particular test system in Ref.\ \citenum{Nishio2019} are variational. All others leave the orbitals, CI vectors, or both constrained indirectly by orbital localization schemes in one way or another. When the orbitals are variationally optimized,\cite{Kim2015,Nishio2019} the authors tend to utilize a standard CASSCF orbital-optimization algorithm, which is mature\cite{Roos1980,Helgaker2000} with many acceleration schemes and approximations to the general second-order approach having been explored.\cite{Werner1980,Werner1985,Chaban1997,Schmidt1998,Ghosh2008,Yanai2009,Kreplin2019}

LASSCF and vLASSCF differ from previously-explored methods in that the orbital optimization protocol is based on a modified DMET algorithm, meaning that the computational cost of not only the CI vector calculation, but also of the electron repulsion integral (ERI) calculation, storage, and handling involved in optimizing the orbitals, is reduced by an embedding formalism. In the following sections, in addition to presenting the vLASSCF extension to LASSCF, we will also show both formally and numerically that the LASSCF protocol of splitting a single large MC-SCF problem into several small coupled MC-SCF problems suppresses the prefactor of the quintic-scaling [$O(M_{\mathrm{tot}}^5)$, where $M_{\mathrm{tot}}$ is the total number of AOs] step required by general second-order CASSCF orbital-optimization algorithms by a factor of (at minimum, in our current implementation) $M_A/M_{\mathrm{tot}}$, where $M_A$ is the total number of active orbitals (in all subspaces combined). This corresponds to effectively reducing the scaling to quartic [$O(M_{\mathrm{tot}}^4)$] in realistic applications.

The rest of this paper is organized as follows. Section \ref{sec:notations} summarizes notational conventions. In Sec.\ \ref{sec:theory_lasscf}, we summarize LASSCF as implemented in Ref.\ \citenum{Hermes2019} and explain its non-variationality. We describe the extension to vLASSCF theoretically in Sec.\ \ref{sec:theory_vlasscf} and analyze the computational cost scaling of orbital optimization in vLASSCF compared to CASSCF formally in Sec.\ \ref{sec:theory_cost_formal}. We present results of test calculations in Sec.\ \ref{sec:results_discussion}, which demonstrate numerically the equivalence of vLASSCF and CASSCF in the appropriate limit, the superior guess stability of vLASSCF compared to CASSCF, and the lower cost scaling of vLASSCF compared to CASSCF. 

\section{Theory \label{sec:theory}}

\subsection{Notational conventions \label{sec:notations}}

\begin{figure}
    \centering
    \includegraphics[width=0.9\textwidth]{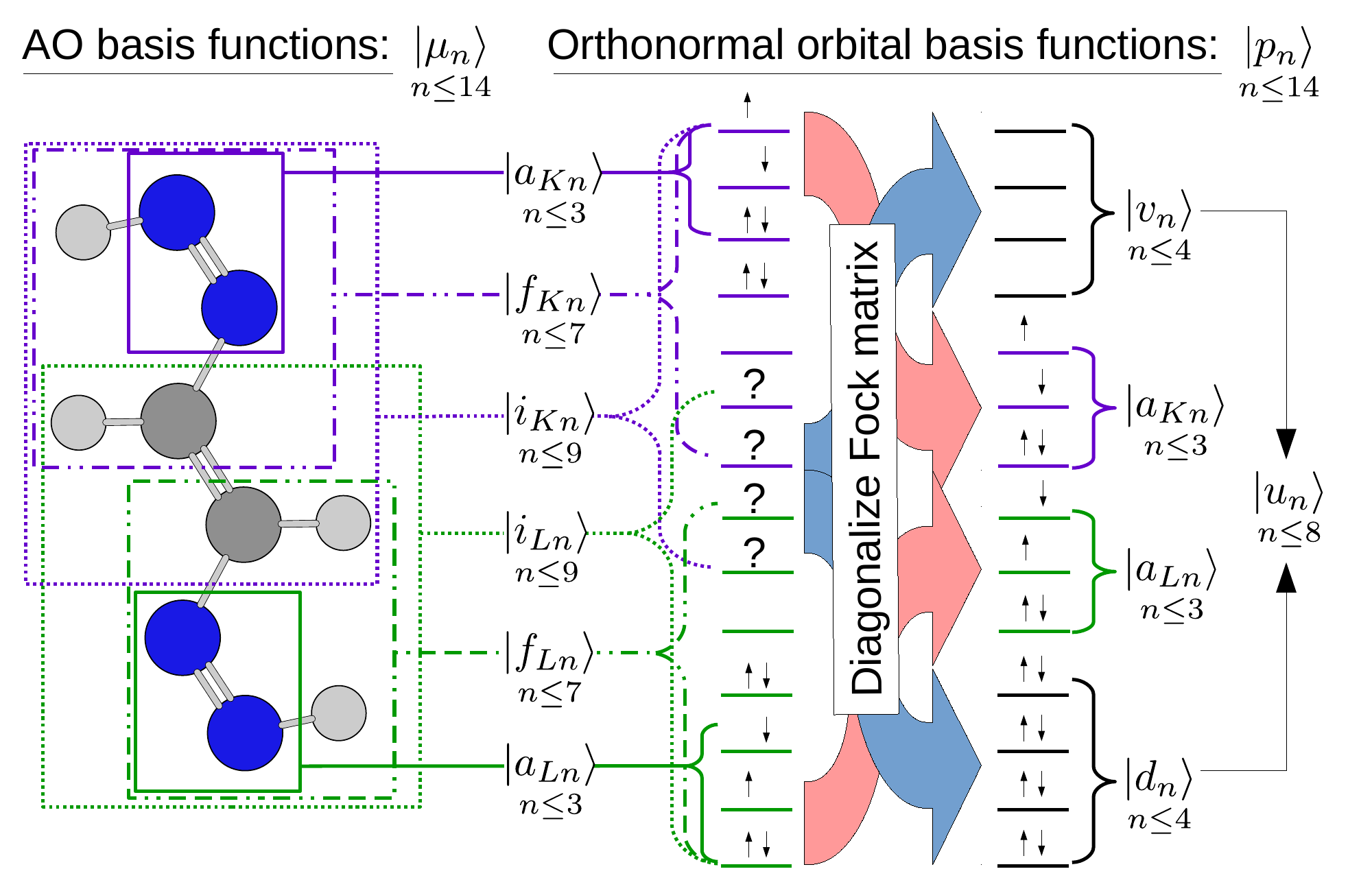}
    \caption{Schematic of the use of 13 types of orbital indices used in this work to describe LASSCF and vLASSCF, along with the limits of the integer subindex $n=1,2,3,\ldots$ for the depicted example. A molecule is depicted on the left, and in this fictional example it is described by 16 electrons occupying 14 AOs ($\ket{\mu_n}$) (this is not intended to be realistic). Those AOs are orthogonalized and split into two fragments, labeled $K$ (purple) and $L$ (green), each of which is characterized by a an average of 8 electrons occupying 7 fragment orbitals ($\ket{f_{Kn}}$), with a corresponding 10-electron, 9-orbital impurity ($\ket{i_{Kn}}$) subspace; each fragment also contains a 4-electron, 3-orbital active ($\ket{a_{Kn}}$) subspace. The orthonormal orbital basis ($\ket{p_n}$) containing active, fragment, and impurity orbitals for both fragments is depicted in the center in a form that resembles a traditional ``energy level diagram'' (although orbital energies in this basis are undefined). Question marks indicate that an orbital cannot be assigned an electron occupancy because the density matrix has off-diagonal elements in that row and column. The energy level diagram for the canonical molecular orbital (MO) basis, exposing inactive ($\ket{d_n}$), virtual ($\ket{v_n}$), and unactive ($\ket{u_n}$) orbitals, is depicted on the right. \label{fig:orbital_diagram}}
\end{figure}

Orbitals are generally referred to using a compound index of the type $s_{Kn}$, where $s$ describes a type of orbital space, $K,L,\ldots$ index different fragments of a large molecule, and $n=1,2,\ldots$ is an integer distinguishing multiple indices in the same space. Any index without a fragment subindex pertains to the entire molecule (i.e., the range of $a_1$ spans all active orbitals in an entire molecule; the range of $a_{K1}$ spans only the active orbitals associated with the $K$th fragment). The symbols $f$, $i$, and $a$ respectively indicate fragment, impurity, and active orbital spaces and may have fragment subindices. The symbols $d$, $v$, $u$, $p$, and $\mu$ respectively refer to doubly-occupied (inactive), virtual (unoccupied), unactive (either inactive or virtual), general molecular, and atomic orbital spaces, which in this work never have fragment subindices and always pertain to the whole molecule. A fictitious example of how these symbols and indices might be used is depicted in Fig.\ \ref{fig:orbital_diagram}. Reference \citenum{Hermes2019} also contains extensive discussions of relationships among several of these subspaces.

An orbital symbol with the literal letter ``$n$'' as a subindex is occasionally used outside of programmable equations as shorthand for ``any and all orbitals of this particular type.'' If the orbital symbol also has pointed brackets around it, for example $\{\ket{p_n}\}$, it indicates the vector space spanned by all orbitals described by that kind of index. An orbital index with an arrow over it, such as $\vec{a}_{K1}$, denotes a configuration state function (CSF) or determinant in the Fock space of $\{\ket{a_{Kn}}\}$. The numerical subindex of the CSF/determinant index distinguishes multiple distinct CSF/determinant indices in the same Fock space.

We consider derivatives of the electronic energy (as an expectation value of the molecular Hamiltonian) with respect to an infinitesimal transformation between orbitals and states. The step vector is written as $\vecX$ with orbital elements $\vecX^{p_1}_{p_2}$ and CSF/determinant (i.e., CI) elements $\vecX_{\vec{a}_{K1}}$. The Taylor expansion of the energy ($E$) in these coordinates is
\begin{eqnarray}
    E &=& \Evec{0} + \frac{1}{2}\Emat{1}^{p_1}_{p_2}\vecX^{p_1}_{p_2} + \sum_K \Evec{1}_{\vec{a}_{K1}}\vecX_{\vec{a}_{K1}} + \frac{1}{2}\sum_K \{\Evec{2}_{\vec{a}_{K1}}\}^{p_1}_{p_2}\vecX_{\vec{a}_{K1}}\vecX^{p_1}_{p_2} \nonumber \\ && + \frac{1}{8}\Emat{2}^{p_1p_3}_{p_2p_4}\vecX^{p_1}_{p_2}\vecX^{p_3}_{p_4} + \frac{1}{2}\sum_{K,L} \Evec{2}_{\vec{a}_{K1}\vec{a}_{L1}}\vecX_{\vec{a}_{K1}}\vecX_{\vec{a}_{L1}} + \ldots  \label{eq:taylor}
\end{eqnarray}
where we implicitly sum over repeated orbital and CSF indices. Equation (\ref{eq:taylor}) displays the symbols for first- and second-order derivatives (gradient and Hessian) of the energy with respect to orbital and CI transformations. The matrix-vector product of the Hessian ($\Evec{2}$) with the step vector ($\vecX$) is itself indexed analogously to the gradient ($\Evec{1}$) and $\vecX$: $\EmatX{2}^{p_1}_{p_2}$ for the orbitals and $\EmatX{2}_{\vec{a}_{K1}}$ for the CI degrees of freedom. The step vector, gradient, Hessian, and Hessian-vector product are all antisymmetric with respect to permutation of any orbital index column: $\vecX^{p_1}_{p_2}=-\vecX^{p_2}_{p_1}$, etc., for which reason a factor of \sfrac{1}{2} is associated with every factor of $\vecX^{p_1}_{p_2}$ to cancel the implicit double-counting.

\subsection{LASSCF \label{sec:theory_lasscf}}

\begin{figure}
\begin{tikzpicture}[node distance=1cm]
\node (start) [startstop] {Start};
\node (input) [io, below of=start] {Guess orbs; fragments};
\node (decconv) [decision, below of=input, yshift=-1cm, text width=1.6cm, aspect=1] {$E$, $\gamma$, $\ket{f_{Kn}}$ fixed point?};
\node (stop) [startstop, left of=decconv, xshift=-1.5cm] {Stop};
\node (profrag) [process, below of=decconv, yshift=-1.5cm, text width=1.5cm] {Shift $\ket{f_{Kn}}$ [Eq.\ (\ref{eq:deltaKL})]};
\node (proKeq0) [process, below of=profrag, yshift=-0.3cm, text width=1.5cm] {$K=0$};
\node (decKloop) [decision, below of=proKeq0, yshift=-1cm] {$K<n_{\mathrm{frag}}$?};
\node (proSchmidt) [process, below of=decKloop, yshift=-1cm] {$\ket{f_{Kn}}\to\ket{i_{Kn}}$ [Eq.\ (\ref{eq:schmidt_add})]};
\node (proimpham) [process, below of=proSchmidt, yshift=-0.5cm] {$\ket{i_{Kn}}\to\hat{H}_{I_K}$ [Eq.\ (\ref{eq:HIK})]};
\node (decscf1) [decision, below of=proimpham, yshift=-1.3cm, text width=1.5cm] {$\Emat{1}^{i_{K1}}_{a_{K1}}=\Evec{1}_{\vec{a}_{K1}}=0$?};
\node (proscf1) [process, below of=decscf1, xshift=1.6cm, yshift=-0.5cm, text width=1cm] {SCF cycle};
\node (proKit) [process, left of=proimpham, xshift=-1.5cm, yshift=0.75cm, text width=1.2cm] {$K\mathrel{+}=1$};
\node (proorth) [process, right of=decKloop, xshift=2cm] {Orthogonalize $\ket{a_{Kn}}$ [Eq.\ (\ref{eq:deltaKL})]};
\node (decscf2) [decision, right of=decconv, xshift=2cm] {$\Emat{1}^{v_1}_{d_1}=0$?};
\node (proscf2) [process, above of=decscf2, xshift=1.6cm, yshift=0.5cm, text width=1cm] {SCF cycle};
\draw [arrow] (start) -- (input);
\draw [arrow] (input) -- (decconv);
\draw [arrow] (decconv) -- node[anchor=east] {no} (profrag);
\draw [arrow] (decconv) -- node[anchor=south] {yes} (stop);
\draw [arrow] (profrag) -- (proKeq0);
\draw [arrow] (proKeq0) -- (decKloop);
\draw [arrow] (decKloop) -- node[anchor=east] {yes} (proSchmidt);
\draw [arrow] (decKloop) -- node[anchor=south] {no} (proorth);
\draw [arrow] (proSchmidt) -- (proimpham);
\draw [arrow] (proimpham) -- (decscf1);
\draw [arrow] (decscf1) |- node[anchor=east] {no} (proscf1);
\draw [arrow] (proscf1) |- (decscf1);
\draw [arrow] (decscf1) -| node[anchor=north] {yes} (proKit);
\draw [arrow] (proKit) |- (decKloop);
\draw [arrow] (proorth) -- (decscf2);
\draw [arrow] (decscf2) |- node[anchor=east] {no} (proscf2);
\draw [arrow] (proscf2) |- (decscf2);
\draw [arrow] (decscf2) -- node[anchor=south] {yes} (decconv); 
\end{tikzpicture}
\caption{Simplified flowchart of the LASSCF algorithm as implemented in Ref.\ \citenum{Hermes2019}.\label{fig:oldLASSCF}}
\end{figure}
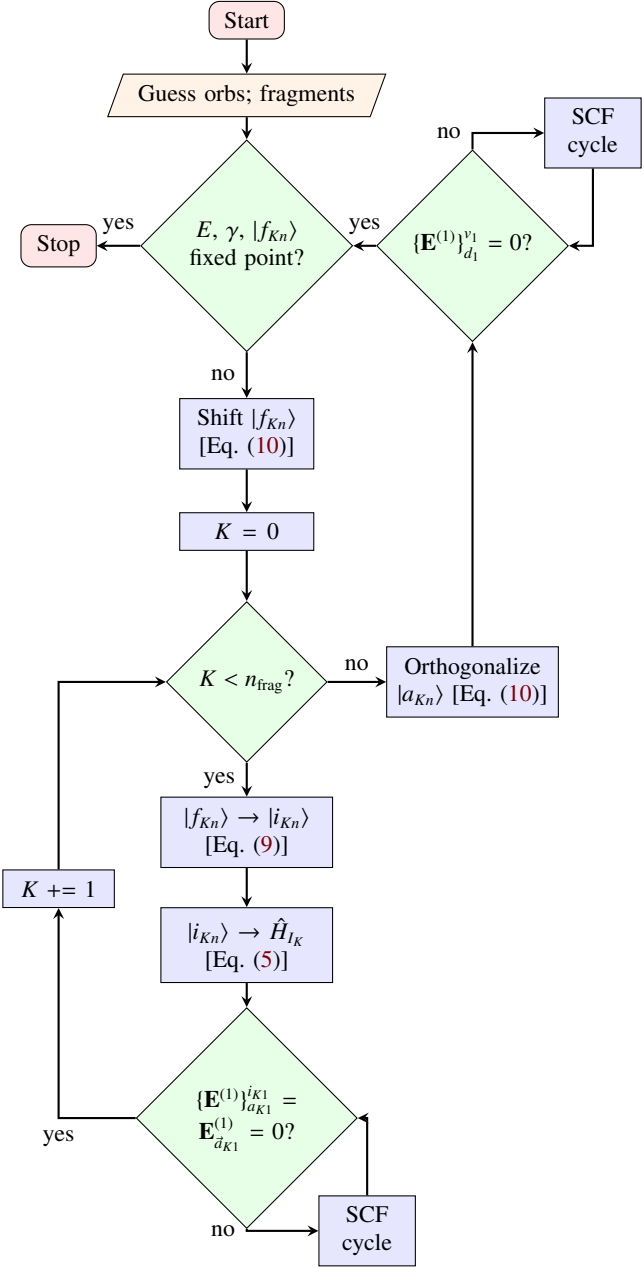

The localized active space wave function and energy are defined as
\begin{eqnarray}
    \ket{\mathrm{LAS}} &=& \left(\bigotimes_K \ket{\Psi_{A_K}}\right)\otimes \ket{\Phi_D},
    \label{eq:lasket}
    \\
    E_{\mathrm{LAS}} &=& \braket{\mathrm{LAS}|\hat{H}|\mathrm{LAS}},
    \label{eq:ELAS}
\end{eqnarray}
where $\hat{H}$ is the standard molecular Hamiltonian, $\ket{\Psi_{A_K}}$ is a general correlated state describing $N_{A_K}$ electrons occupying the $M_{A_K}$ active orbitals ($\ket{a_{Kn}}$) of the $K$th fragment, and $\ket{\Phi_D}$ is a single determinant of doubly-occupied inactive orbitals ($\ket{d_n}$). The LASSCF method obtains $\ket{\mathrm{LAS}}$ in several overlapping sets of impurity orbitals ($\ket{i_{Kn}}$):
\begin{eqnarray}
    \ket{\Psi_{A_K}} &=& \argmin_{\ket{\Psi_{A_K}}} \left(\min_{\ket{\Phi_{I_K}}} \bra{\Phi_{I_K}}\otimes\braket{\Psi_{A_K}|\hat{H}_{I_K}|\Psi_{A_K}}\otimes\ket{\Phi_{I_K}}\right). \label{eq:imp_casscf}
\end{eqnarray}
where $\ket{\Phi_{I_K}}$ is a single determinant of doubly-occupied linear combinations of impurity orbitals, and the impurity Hamiltonian ($\hat{H}_{I_K}$) is
\begin{eqnarray}
    \hat{H}_{I_K} &=& \left(h^{i_{K1}}_{i_{K2}} 
    +\{v_\sigma^{(jk)}\}^{i_{K1}}_{i_{K2}}
    -\{v_\sigma^{(\mathrm{self})}\}^{i_{K1}}_{i_{K2}}\right)
    \crop{i_{K1}\sigma}\anop{i_{K2}\sigma}
    \nonumber\\ && +\frac{1}{2}
    g^{i_{K1}i_{K3}}_{i_{K2}i_{K4}}
    \crop{i_{K1}\sigma}\crop{i_{K3}\tau}\anop{i_{K4}\tau}\anop{i_{K2}\sigma},
    \label{eq:HIK}
\end{eqnarray}
omitting an irrelevant constant, where
\begin{eqnarray}
    \{v_\sigma^{(jk)}\}^{\mu_1}_{\mu_2} &=& g^{\mu_1\mu_3}_{\mu_2\mu_4} D^{\mu_3}_{\mu_4} - g^{\mu_1\mu_3}_{\mu_4\mu_2} \{\gamma_\sigma\}^{\mu_3}_{\mu_4}, \label{eq:vjk}
    \\
    \{v_\sigma^{(\mathrm{self})}\}^{i_{K1}}_{i_{K2}} &=& g^{i_{K1}i_{K3}}_{i_{K2}i_{K4}} D^{i_{K3}}_{i_{K4}} - g^{i_{K1}i_{K3}}_{i_{K4}i_{K2}} \{\gamma_\sigma\}^{i_{K3}}_{i_{K4}}, \label{eq:vself}
\end{eqnarray}
and where $\crop{i_{Kn}\sigma}$ ($\anop{i_{Kn}\sigma}$) creates (annihilates) an electron at the $i_{Kn}$th orbital with spin $\sigma$, $h$ and $g$ are respectively the one- and two-electron molecular Hamiltonian matrix elements, $D$ and $\gamma_\sigma$ are respectively the spin-summed and spin-separated one-body reduced density matrices (1-RDMs; $D=\gamma_\uparrow+\gamma_\downarrow$), and we again sum over repeated internal indices including the spin indices $\sigma$ and $\tau$ (which take the values $\uparrow$ and $\downarrow$), but not  the fragment subindices $K$ [e.g., $i_{K3}$ in Eq.\ (\ref{eq:vself}) ranges over the impurity orbitals of the $K$th subspace only]. The impurity orbitals are obtained by combining $M_{F_K}$ ``fragment'' orbitals with up to $M_{F_K}$ entangled partner orbitals \emph{via} the Schmidt decomposition [i.e., singular value decomposition (SVD) of the density matrix]:\cite{Schmidt1907,Peschel2012}
\begin{eqnarray}
    \left(1-c^{p_1}_{f_{K1}}c^{f_{K1}}_{p_2}\right)D^{p_2}_{f_{K2}} &=& \sum_{p_3}^{M_{F_K}} u^{p_1}_{p_3} \sigma_{p_3} v^{f_{K2}}_{p_3}, \label{eq:schmidt}
    \\
    \{\ket{f_{Kn}}\} \cup \left\{\ket{p_1}u^{p_1}_{p_3}\right\} &\to& \{\ket{i_{Kn}}\}, \label{eq:schmidt_add}
\end{eqnarray}
where $c^{p_1}_{f_{K1}}$ is an element of a unitary transformation matrix between fragment orbitals and some general set of molecular orbitals, and only the left-singular vectors ($u^{p_1}_{p_3}$) corresponding to nonzero singular values ($\sigma_{p_3}$) are retained. The role of the Schmidt decomposition is to augment the fragment orbitals, which are entangled to the rest of the molecule and cannot be assigned a wave function on their own, with additional degrees of freedom that account for all entanglement and therefore ensure that the impurity space is occupied by an integer number of electrons;\cite{Knizia2012} this allows Eq.\ (\ref{eq:imp_casscf}) to be solved with standard implementations of CASSCF.\cite{Hermes2019} The fragment orbitals themselves are non-overlapping sets of localized orbitals which collectively span the whole AO space and which each must enclose at most exactly one set of active subspace orbitals:
\begin{eqnarray}
    \sum_{f_{L1}} |\braket{a_{K1}|f_{L1}}|^2 &=& \delta_{KL},
    \label{eq:deltaKL}
\end{eqnarray}
The fragment orbitals are chosen to resemble a set of orthogonalized AOs as closely as possible while satisfying Eq.\ (\ref{eq:deltaKL}). In a single cycle of the LASSCF iteration, Eqs.\ (\ref{eq:imp_casscf})--(\ref{eq:deltaKL}) are used for each fragment to obtain an updated guess for the active orbitals and CI vectors. The active orbitals will tend to develop nonzero overlaps across fragments from being optimized asynchronously and must be explicitly orthogonalized once per cycle. They are then frozen along with the CI vectors while the whole-molecule inactive orbitals ($\ket{d_n}$) are optimized. The cycle repeats until fragment orbitals, density matrices, and energies stop changing. Figure \ref{fig:oldLASSCF} presents a schematic of the these steps.

LASSCF will be less computationally costly than CASSCF for large systems for at least two reasons.  First, if a large active space can be split into several small active spaces, then the exponential or factorial cost scaling with respect to the number of orbitals for the determination of the active space CI vector is transformed into linear scaling with respect to the number of active subspaces. Secondly, a lower-scaling number of ERIs are required to optimize the orbitals. LASSCF only requires explicit ERI evaluation in the diagonal blocks of impurity subspaces; the size of the ERI array in cache must have asymptotic linear dependence on the size of the molecule if the number of fragments increases with the system size.  This is discussed in more detail in Sec.\ \ref{sec:theory_cost_formal}.

There is a cost to this computational efficiency. Equations (\ref{eq:imp_casscf})--(\ref{eq:deltaKL}) describe an energy minimization protocol, thus providing an upper bound to the FCI energy; but this is a constrained energy minimization because the $K$th set of active orbitals are only allowed to relax within the $K$th impurity space. That is, at convergence, the LAS wave function satisfies,
\begin{equation}
    \Evec{1}_{\vec{a}_{K1}} = \Emat{1}^{a_{K1}}_{i_{K1}} = \Emat{1}^{v_1}_{d_1} = 0, \label{eq:las_relaxed}
\end{equation}
for all indices, but generally does \emph{not} satisfy
\begin{equation}
    \Emat{1}^{a_{K1}}_{a_{L1}} = \Emat{1}^{a_{K1}}_{u_1} = 0, \label{eq:relaxme}
\end{equation}
where, as a reminder, $\Evec{1}_{\vec{a}_{K1}}$ is the first derivative of the LAS energy with respect to a shift of the wave function in the $K$th active subspace to a determinant or CSF labeled $\vec{a}_{K1}$, and $\Emat{1}^{p_1}_{p_2}$ is the first derivative with respect to rotation of the orbital $\ket{p_1}$ with $\ket{p_2}$.

Constrained energy minimizations are not necessarily inherently problematic for reproducibility and robustness. But in this case, the reduction of the whole molecule's orbital space into an ``impurity'' subspace, which describes the only orbitals the active space is allowed to explore, \emph{itself depends on the active orbitals}. Comparison of Eqs.\ (\ref{eq:las_relaxed}), (\ref{eq:schmidt}), and (\ref{eq:deltaKL}) exposes this self-referentialism: the impurity orbitals constrain the active orbitals, the fragment orbitals define the impurity orbitals, and the active orbitals constrain the fragment orbitals. The ultimate origin of the constraint is simply the initial guesses provided by the user for the active orbitals, the fragment orbitals, or both. 

\subsection{vLASSCF \label{sec:theory_vlasscf}}

Here, we describe an extension to the LASSCF method which renders it truly variational in the sense that the Hellmann-Feynman theorem\cite{Hellmann1937,Feynman1939,Pulay1987} applies:
\begin{equation}
    \Evec{1}_{\vec{a}_{K1}} = \Emat{1}^{p_1}_{p_2} = 0.
    \label{eq:vLASSCF_conv}
\end{equation}
For clarity, we refer to this method as ``variational LASSCF'' or vLASSCF; throughout the rest of this work, ``LASSCF'' refers strictly to the method and implementation described in Ref.\ \citenum{Hermes2019}. The algorithm is modified in two places; once to address each class of unoptimized orbital rotations indicated in Eq.\ (\ref{eq:relaxme}). The first modification is detailed in Sec.\ \ref{sec:lasci_scf} and the second in Sec.\ \ref{sec:gradhess_imp} below. The overall vLASSCF algorithm is depicted in Fig. \ref{fig:vLASSCF}. 

\begin{figure}
\begin{tikzpicture}[node distance=1cm]
\node (start) [startstop] {Start};
\node (input) [io, below of=start] {Guess orbs; fragments};
\node (decconv) [vdecision, below of=input, yshift=-1cm, text width=1.3cm, aspect=1] {\textcolor{red}{$\Emat{1}^{p_1}_{p_2}=\Evec{1}_{\vec{a}_{K1}}=0$?}};
\node (stop) [startstop, left of=decconv, xshift=-1.5cm] {Stop};
\node (profrag) [process, below of=decconv, yshift=-1.5cm, text width=1.5cm] {Shift $\ket{f_{Kn}}$ [Eq.\ (\ref{eq:deltaKL})]};
\node (proKeq0) [process, below of=profrag, yshift=-0.3cm, text width=1.5cm] {$K=0$};
\node (decKloop) [decision, below of=proKeq0, yshift=-1cm] {$K<n_{\mathrm{frag}}$?};
\node (progradorb) [vprocess, below of=decKloop, yshift=-1cm] {\textcolor{red}{$\mathbf{E}^{(1)}\to\ket{f_{Kn}}$ [Eq.\ (\ref{eq:gradorbs_add})]}};
\node (proSchmidt) [process, right of=progradorb, xshift=2cm] {$\ket{f_{Kn}}\to\ket{i_{Kn}}$ [Eq.\ (\ref{eq:schmidt_add})]};
\node (prohessorb) [vprocess, below of=proSchmidt, yshift=-0.5cm] {\textcolor{red}{$\mathbf{E}^{(2)}\to\ket{i_{Kn}}$ [Eq.\ (\ref{eq:hessorbs_add})]}};
\node (proimpham) [process, below of=progradorb, yshift=-0.5cm] {$\ket{i_{Kn}}\to \hat{H}_{I_K}$ [Eq.\ (\ref{eq:HIK})]};
\node (decscf1) [decision, below of=proimpham, yshift=-1.3cm, text width=1.5cm] {$\Emat{1}^{i_{K1}}_{a_{K1}}=\Evec{1}_{\vec{a}_{K1}}=0$?};
\node (proscf1) [process, below of=decscf1, xshift=1.6cm, yshift=-0.5cm, text width=1cm] {SCF cycle};
\node (proKit) [process, left of=proimpham, xshift=-1.5cm, yshift=0.75cm, text width=1.2cm] {$K\mathrel{+}=1$};
\node (proorth) [process, right of=decKloop, xshift=2cm] {Orthogonalize $\ket{a_{Kn}}$ [Eq.\ (\ref{eq:deltaKL})]};
\node (decscf2) [vdecision, right of=decconv, xshift=2cm, align=left] {\parbox{1.0cm}{\centering $\Emat{1}^{v_1}_{d_1}=\textcolor{red}{\Emat{1}^{a_{K1}}_{a_{L1}}=}$\\$\textcolor{red}{\Evec{1}_{\vec{a}_{K1}}=}\ 0$?}};
\node (proscf2) [process, above of=decscf2, xshift=1.6cm, yshift=0.7cm, text width=1cm] {SCF cycle};
\draw [arrow] (start) -- (input);
\draw [arrow] (input) -- (decconv);
\draw [arrow] (decconv) -- node[anchor=east] {no} (profrag);
\draw [arrow] (decconv) -- node[anchor=south] {yes} (stop);
\draw [arrow] (profrag) -- (proKeq0);
\draw [arrow] (proKeq0) -- (decKloop);
\draw [arrow] (decKloop) -- node[anchor=east] {yes} (progradorb);
\draw [arrow] (progradorb) -- (proSchmidt);
\draw [arrow] (decKloop) -- node[anchor=south] {no} (proorth);
\draw [arrow] (proSchmidt) -- (prohessorb);
\draw [arrow] (prohessorb) -- (proimpham);
\draw [arrow] (proimpham) -- (decscf1);
\draw [arrow] (decscf1) |- node[anchor=east] {no} (proscf1);
\draw [arrow] (proscf1) |- (decscf1);
\draw [arrow] (decscf1) -| node[anchor=north] {yes} (proKit);
\draw [arrow] (proKit) |- (decKloop);
\draw [arrow] (decscf2) |- node[anchor=east] {no} (proscf2);
\draw [arrow] (proscf2) |- (decscf2);
\draw [arrow] (decscf2) -- node[anchor=south] {yes} (decconv);
\draw [arrow] (proorth) -- (decscf2);
\end{tikzpicture}
\caption{Simplified flowchart of the vLASSCF algorithm. Differences from LASSCF are indicated by red text and thick red node boxes. The highlighted decision node (diamond) at the center top describes the convergence criteria of a fully variational method. The highlighted decision node at the top right corresponds to the modifications described in Sec.\ \ref{sec:lasci_scf}, and the highlighted process nodes (rectangles) between the center and bottom decision nodes correspond to the modifications described in Sec.\ \ref{sec:gradhess_imp}.\label{fig:vLASSCF}}
\end{figure}
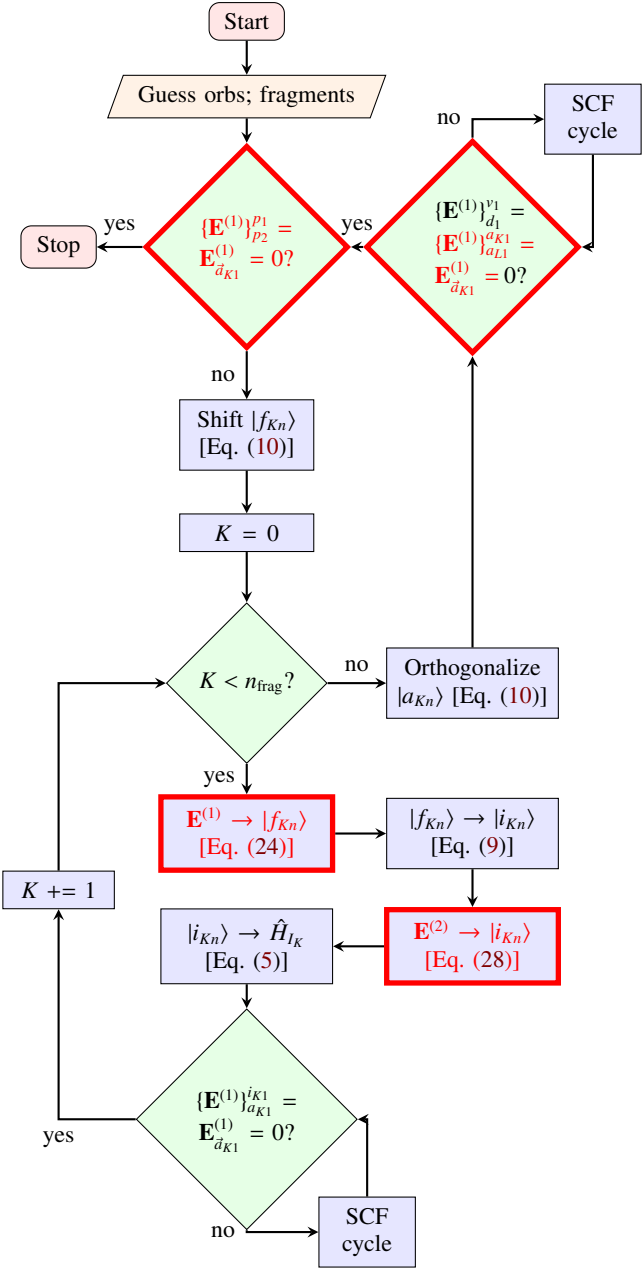 

\subsubsection{Inter-subspace active orbital optimization\label{sec:lasci_scf}}

At the step of the LASSCF cycle where the whole-molecule inactive orbitals are optimized with the active subspaces fixed (top right decision node in Figs.\ \ref{fig:oldLASSCF} and \ref{fig:vLASSCF}), vLASSCF additionally minimizes the energy with respect to rotations between two or more active subspaces and their CI vectors:
\begin{equation}
    \Evec{1}_{\vec{a}_{K1}} = \Emat{1}^{a_{K1}}_{a_{L1}} = \Emat{1}^{v_1}_{d_1} = 0, \label{eq:lasci_scf}
\end{equation}
with the overall shape of the active superspace fixed at that determined by the previous round (loop over $K$ in the lower half of Figs.\ \ref{fig:oldLASSCF} and \ref{fig:vLASSCF}) of impurity calculations:
\begin{equation}
    \vecX^{a_{K1}}_{d_1} = \vecX^{v_1}_{a_{K1}} = 0, \label{eq:lasci_frozen_active}
\end{equation}
where, as a reminder, $\vecX^{p_1}_{p_2}=-\vecX^{p_2}_{p_1}$ is the displacement along the rotation coordinate between the $p_1$th and $p_2$th orbitals.

The first derivatives in Eq.\ (\ref{eq:lasci_scf}) are, explicitly,
\begin{eqnarray}
    \Evec{1}_{\vec{a}_{K1}} &=& \braket{\vec{a}_{K1}|\hat{Q}_{\Psi_{A_K}}\hat{H}_{A_K}|\Psi_{A_K}} + \mathrm{h.c.},
    \\
    \Emat{1}^{a_{K1}}_{a_{L1}} &=& \Fmat{1}^{a_{K1}}_{a_{L1}} - \Fmat{1}^{a_{L1}}_{a_{K1}},
    \\
    \Emat{1}^{v_1}_{d_1} &=& 2\left(h^{v_1}_{d_1} + \{v_\sigma^{(jk)}\}^{v_1}_{d_1}\right),
\end{eqnarray}
where $\hat{H}_{A_K}$ is analogous to $\hat{H}_{I_K}$ given by Eqs.\ (\ref{eq:HIK})--(\ref{eq:vself}), but with all impurity orbital ($\ket{i_{Kn}}$) indices replaced with active orbital ($\ket{a_{Kn}}$) indices; ``h.c.'' means Hermitian conjugate; and
\begin{eqnarray}
    \hat{Q}_{\Psi_{A_K}} &=& 1-\ket{\Psi_{A_K}}\bra{\Psi_{A_K}},
    \\
    \Fmat{1}^{a_{K1}}_{a_{L1}} &=& \left(h^{a_{K1}}_{a_{L2}} + \{v_\sigma^{(jk)}\}^{a_{K1}}_{a_{L2}}\right) \{\gamma_\sigma\}^{a_{L1}}_{a_{L2}} + g^{a_{K1}a_{L3}}_{a_{L2}a_{L4}} \lambda^{a_{L1}a_{L3}}_{a_{L2}a_{L4}}, \label{eq:Fmat1}
\end{eqnarray}
where $\lambda$ is the cumulant\cite{Kutzelnigg1999} of the two-body reduced density matrix (2-RDM),
\begin{eqnarray}
    \lambda^{a_{K1}a_{K3}}_{a_{K2}a_{K4}} &=& \braket{\mathrm{LAS}|\crop{a_{K1}\sigma}\crop{a_{K3}\tau}\anop{a_{K4}\tau}\anop{a_{K2}\sigma}|\mathrm{LAS}}
    \nonumber\\ && - D^{a_{K1}}_{a_{K2}} D^{a_{K3}}_{a_{K4}} + \{\gamma_\sigma\}^{a_{K1}}_{a_{K4}} \{\gamma_\sigma\}^{a_{K3}}_{a_{K2}}, \label{eq:lambda}
\end{eqnarray}
The solution of Eq.\ (\ref{eq:lasci_scf}) is obtained by a fully second-order algorithm; repeatedly solving
\begin{eqnarray}
    \mathbf{E}^{(1)} + \mathbf{E}^{(2)}\vecX &=& \vec{0},
\end{eqnarray}
for $\vecX$, which contains all orbital- and CI-transformation variables. The Hessian-vector products, $\mathbf{E}^{(2)}\vecX$, are tabulated in Sec.\ 1.3 of the SI.

Evaluation of $\hat{H}_{A_K}$ and $\Fmat{1}^{a_{K1}}_{a_{L1}}$ requires the HF-like effective potential, $v_\sigma^{(jk)}$, as well as ERIs with orbital indices spanning up to two active subspaces in the last term of Eq.\ (\ref{eq:Fmat1}). In our current implementation, we evaluate ERIs spanning all active orbitals in all subspaces (i.e., $g^{a_{K1}a_{M1}}_{a_{L1}a_{N1}}$) once per LASSCF macrocycle, which allows quick re-evaluation of $g^{a_{K1}a_{L3}}_{a_{L2}a_{L4}}$ after mixing the active subspaces without requiring another reference to AO-basis ERIs. This is not strictly necessary and incurs an overall quintic cost scaling for the method with respect to the total number of active orbitals [$O(M_A^5)$], but it remains a very fast step for all test cases here explored. We return to this issue in Sec.\ \ref{sec:theory_cost_formal} below. 

\subsubsection{Generation of impurity subspaces that allow full active-orbital relaxation\label{sec:gradhess_imp}}

The other difference between the LASSCF and vLASSCF algorithms is in the construction of the impurity orbitals ($\ket{i_{Kn}}$, between the center and bottom decision nodes of Figs.\ \ref{fig:oldLASSCF} and \ref{fig:vLASSCF})  At this point in the calculation, the energy is stationary with respect to all variables except for the shape of the active superspace (that is, all active orbitals for all fragments collectively, $\cup_K \{\ket{a_{Kn}}\}$). We rely on the impurity calculations [Eq.\ (\ref{eq:imp_casscf}) for each $K$] to optimize the shape of the active superspace; that is, to minimize the energy by rotating active orbitals with inactive and external orbitals. However, as discussed in Sec.\ \ref{sec:theory_lasscf}, the space within which this rotation occurs for the $K$th fragment (the impurity space, $\{\ket{i_{Kn}}\}$) is not complete in LASSCF, and there is no guarantee that the active orbitals ever minimize the LAS energy [i.e., satisfy Eq.\ (\ref{eq:vLASSCF_conv})]. 

In vLASSCF, we solve this problem by adding orbitals to the impurity subspace which are constructed using the \emph{orbital rotation gradient and Hessian}. The gradient is a two-index object describing pairs of orbitals and is therefore amenable to SVD; if there are $M_A\equiv\sum_K M_{A_K}$ active orbitals in the whole molecule, then at most $M_A$ linear combinations of unactive orbitals characterize the \emph{entire} gradient, and repeated energy minimizations in impurity subspaces which contain one gradient-coupled unactive partner for each active orbital are guaranteed to eventually satisfy Eq.\ (\ref{eq:vLASSCF_conv}). Since these gradient-coupled unactive orbitals may be entangled to other orbitals, we simply add them to the fragment subspace,
\begin{eqnarray}
    \left(1-c^{u_1}_{f_{K1}}c^{f_{K1}}_{u_2}\right)\Emat{1}^{u_2}_{a_{K1}} &=& \sum_{p_1}^{M_{A_K}} u^{u_1}_{p_1}\sigma_{p_1}\{v^{a_{K1}}_{p_1}\}^*, \label{eq:gradorbs}
    \\
    \{\ket{f_{Kn}}\}\cup\left\{\ket{u_1}u^{u_1}_{p_1}\right\} &\to& \{\ket{f_{Kn}}\}, \label{eq:gradorbs_add}
\end{eqnarray}
and the Schmidt decomposition [Eqs.\ (\ref{eq:schmidt})-(\ref{eq:schmidt_add})] will compensate for any entanglement to the rest of the molecule by adding all necessary degrees of freedom to the impurity. The projection on the left-hand side of Eq.\ (\ref{eq:gradorbs}) simply ensures that the fragment orbitals remain orthonormal. The gradient in Eq.\ (\ref{eq:gradorbs}) is evaluated as
\begin{eqnarray}
    \Emat{1}^{u_1}_{a_{K1}} &=& \left(h^{u_1}_{a_{K2}} + \{v_\sigma^{(jk)}\}^{u_1}_{a_{K2}}\right)\{\gamma_\sigma\}^{a_{K2}}_{a_{K1}} 
    \nonumber \\ && -
    \left(h^{a_{K1}}_{u_2} + \{v_\sigma^{(jk)}\}^{a_{K1}}_{u_2}\right)\{\gamma_\sigma\}^{u_2}_{u_1}
    \nonumber \\ &&
    + g^{u_1a_{K3}}_{a_{K2}a_{K4}}\lambda^{a_{K1}a_{K3}}_{a_{K2}a_{K4}}, \label{eq:E1ua}
\end{eqnarray}
which requires only ERIs with three active indices (in the same subspace) and one unactive index, as well as HF-like effective potential terms $v_\sigma^{(jk)}$.

The maximum number of impurity orbitals taking into account both Eq.\ (\ref{eq:gradorbs}) and Eq.\ (\ref{eq:schmidt}) is $2M_{F_K}$ (unactive fragment orbitals, including those constructed from SVD of the gradient as described above, with one entangled partner each) plus $M_{A_K}$ (active orbitals, which are not entangled outside of the fragment\cite{Hermes2019} and so cannot generate additional impurity orbitals \emph{via} the Schmidt decomposition). However, LASSCF inherits\cite{Hermes2019,Pandharkar2019} from DMET an instability with respect to the size of the AO basis set. The entanglement described by the nonzero singular values in Eq.\ (\ref{eq:schmidt}) is based on doubly-occupied inactive ($\ket{d_n}$) orbitals, which as the size of the AO basis grows represents a smaller and smaller part of the full AO space.\cite{Wouters2017} Therefore, impurity spaces tend to lack high-lying, diffuse virtual orbitals in DMET and LASSCF calculations and the total number of impurity orbitals falls short of $2M_{F_K}+M_{A_K}$ when large AO basis sets are used.  This will tend to make convergence of the vLASSCF macroiteration slower, since each impurity calculation can access a relatively smaller number of AOs.

To cure this deficiency, we further augment the impurity space with additional orbitals obtained by manipulation of the Hessian, $\mathbf{E}^{(2)}$, which at this point is the leading-order nonzero term in the Taylor series expansion of the LAS energy [Eq.\ (\ref{eq:taylor})] which couples the impurity orbitals to the rest of the molecule. We construct an approximate orbital-optimization step vector within the impurity subspace ($\vecX$, with elements $\vecX^{i_{K1}}_{i_{K2}}$) and then utilize the Hessian-vector product, $\mathbf{E}^{(2)}\vecX$, analogously to the gradient in Eq.\ (\ref{eq:gradorbs}),
\begin{eqnarray}
    \left(1-c^{d_1}_{i_{K1}}c^{i_{K1}}_{d_2}\right)\EmatX{2}^{d_2}_{i_{K1}} &=& \sum_{p_1} u^{d_1}_{p_1}\sigma_{p_1}\{v^{i_{K1}}_{p_1}\}^*, \label{eq:hessorbs_d}
    \\
    \left(1-c^{v_1}_{i_{K1}}c^{i_{K1}}_{v_2}\right)\EmatX{2}^{v_2}_{i_{K1}} &=& \sum_{p_1} u^{v_1}_{p_1}\sigma_{p_1}\{v^{i_{K1}}_{p_1}\}^*, \label{eq:hessorbs_v}
    \\
    \{\ket{i_{Kn}}\}\cup\left\{\ket{u_1}u^{u_1}_{p_1}\right\} &\to& \{\ket{i_{Kn}}\}, \label{eq:hessorbs_add}
\end{eqnarray}
where the singular values and vectors of Eqs.\ (\ref{eq:hessorbs_d}) and (\ref{eq:hessorbs_v}) are concatenated and used to add impurity orbitals in descending order of the magnitude of $\sigma_{p_1}$ until the size of the impurity reaches $2(M_{F_K}+M_{A_K})$, which in principle corresponds to the maximum size of the impurity as given before, plus $M_{A_K}$ additional Hessian-coupled unactive orbitals. However, this step also can compensate for the basis set instability problem discussed above, since $\EmatX{2}^{u_1}_{i_{K1}}$ generally has more than $M_{A_K}$ nonzero singular values. (As a reminder, the unactive orbitals are the union of inactive and virtual orbitals: $\{\ket{u_n}\}=\{\ket{d_n}\}\cup\{\ket{v_n}\}$.) It is somewhat arbitrary that we have chosen to carry out one SVD of the Hessian-vector product for the inactive orbitals and another for the external orbitals [Eq.\ (\ref{eq:hessorbs_d}) and (\ref{eq:hessorbs_v})], instead of augmenting the fragment subspace before Schmidt decomposition as was done for the gradient [Eq.\ (\ref{eq:gradorbs})]; these alternatives are not necessarily equivalent, because, e.g., both $\ket{u_1}$ and $\ket{i_{K1}}$ in $\EmatX{2}^{u_1}_{i_{K1}}$ may contain components outside of $\{\ket{f_{Kn}}\}$, but it is unknown whether this distinction has significant consequences.

The expressions for the Hessian-vector products appearing in Eqs.\ (\ref{eq:hessorbs_d}) and (\ref{eq:hessorbs_v}), including the evaluation of the step vector $\vecX$, are tabulated in Sec.\ 1.5 of the SI. We expect these expressions to be robust to approximation, because we do not need to actually minimize the energy using the Hessian in this step. We therefore explore two protocols for evaluating these Hessian elements in Secs.\ \ref{sec:theory_cost_formal} and \ref{sec:results_discussion} below: in the ``full Hessian'' protocol, the Hessian is evaluated without approximation, whereas in the ``approximate Hessian'' protocol, we drop from the expressions in the SI all of the terms which depend on explicit ERIs other than those already evaluated to calculate the orbital gradient. Without here reproducing the expressions in the SI, the implicated ERIs can be inferred from the types of orbital indices which appear in the unpacked expression for the Hessian-vector product:
\begin{eqnarray}
    \EmatX{2}^{u_1}_{i_{K1}} &=& \Emat{2}^{u_1i_{K2}}_{i_{K1}i_{K3}} \vecX^{i_{K2}}_{i_{K3}}. \label{eq:Hx_unpack}
\end{eqnarray}
The index pattern of the explicit ERIs required for Eqs.\ (\ref{eq:hessorbs_d}) and (\ref{eq:hessorbs_v}) in the ``full Hessian'' protocol is the same as that of the Hessian elements on the right-hand side of Eq.\ (\ref{eq:Hx_unpack}): $g^{u_1i_{K2}}_{i_{K1}i_{K3}}$. This syllogism works whenever the unique orbital ranges indicated by the indices (here $i_{Kn}$ and $u_1$) all represent unentangled subspaces of the molecule as explained further in Sec.\ 1.4 of the SI.

\subsection{Orbital optimization operation cost scaling formal analysis \label{sec:theory_cost_formal}}

In the following analysis, we consider the case of Cholesky decomposition and density fitting\cite{Koch2003} approach to the ERIs and to the effective potentials,
\begin{eqnarray}
    g^{\mu_1\mu_3}_{\mu_2\mu_4} &=& (\mu_1\mu_2|P)\left\{(P|Q)^{-1}\right\}(\mu_3\mu_4|Q)
    \nonumber \\ &=& b^P_{\mu_1\mu_2} b^{P}_{\mu_3\mu_4},
\end{eqnarray}
where $P,Q$ index the auxiliary AOs, $(\mu_1\mu_2|P)$ and $(P|Q)$ are respectively the physical three-center and two-center ERIs, and $b^P_{\mu_1\mu_2}$ are the Cholesky-decomposed three-center ERIs. We also assume that in LASSCF or vLASSCF, as the molecule's size increases, the number of fragments increases: $n_{\mathrm{frag}}\propto M_{\mathrm{tot}}^1$. Note that $n_{\mathrm{frag}}$ is the range of the fragment subindex, $K$. The number of auxiliary AOs is denoted $M_{\mathrm{aux}}$. We have not considered the sparsity of the three-center integrals in terms of distance between the AO centers in this analysis.

The computational hot spots in CASSCF, LASSCF, and vLASSCF can collectively be grouped into three categories:
\begin{enumerate}
    \item Solution of the CI problem in the active (sub)space(s),
    \item Repeated calculation of effective potential $v_\sigma^{(jk)}$ [Eq.\ (\ref{eq:vjk})] during SCF cycles,
    \item Calculation of explicit cached ERIs.
\end{enumerate}
We ignore the first category in this work since the obvious conclusion was presented in Sec.\ \ref{sec:intro} and is the same as for many other formalisms proposed in the literature.\cite{Parker2013,Parker2014,Parker2014a,Kim2015,Jimenez-Hoyos2015,Nishio2019} We explore these remaining two categories in the subsections below.

\subsubsection{Calculation of effective potentials}

The effective potentials $v_\sigma^{(jk)}$ are evaluated in the AO basis, as indicated in Eqs.\ (\ref{eq:vjk}). [In contrast, $v_\sigma^{(\mathrm{self})}$ in Eq.\ (\ref{eq:vself}) is presented in terms of impurity orbitals and is evaluated using cached ERIs.] They can each be separated into a Coulomb ($v^{(j)}$) and exchange ($v_\sigma^{(k)}$) term which are evaluated separately:
\begin{eqnarray}
    \{v_\sigma^{(jk)}\}^{\mu_1}_{\mu_2} &=& \{v^{(j)}\}^{\mu_1}_{\mu_2} - \{v_\sigma^{(k)}\}^{\mu_1}_{\mu_2},
\end{eqnarray}
The evaluation of $v^{(j)}$ using a general density matrix in the AO basis with density fitting is very fast in practice, and consists of two steps,
\begin{eqnarray}
    \rho_P &=& b^P_{\mu_1\mu_2} D^{\mu_1}_{\mu_2},
    \\
    \{v^{(j)}\}^{\mu_1}_{\mu_2} &=& \rho_P b^P_{\mu_1\mu_2},
\end{eqnarray}
both of which have operation cost scaling of $O(M_{\mathrm{aux}}M_{\mathrm{tot}}^2)$. On the other hand, $v_\sigma^{(k)}$ requires a three-index intermediate,
\begin{eqnarray}
    \{\chi_{\sigma}\}^P_{\mu_1\mu_3} &=& b^P_{\mu_1\mu_2}\{\gamma_\sigma\}^{\mu_2}_{\mu_3}, \label{eq:v3body}
    \\
    \{v_\sigma^{(k)}\}^{\mu_1}_{\mu_2} &=& \{\chi_\sigma\}^P_{\mu_1\mu_3}b^P_{\mu_2\mu_3},
\end{eqnarray}
and therefore has operation cost scaling of $O(M_{\mathrm{aux}}M_{\mathrm{tot}}^3)$ (in both steps). All methods here discussed have this repeated step in common.

\subsubsection{Explicit ERIs}

\paragraph{CASSCF ERIs.} CASSCF orbital optimization requires repeated explicit ERI evaluation in the index patterns $g^{p_1a_1}_{p_2a_2}$ as well as $g^{p_1p_2}_{a_1a_2}$ in order to evaluate the orbital-rotation Hessian-vector products.\cite{Roos1980,Helgaker2000,Ghosh2008} In the context of density fitting (at least as implemented in {\sc PySCF}\cite{Sun2018}), these are evaluated as
\begin{eqnarray}
    b^P_{p_1\mu_2} &=& b^P_{\mu_1\mu_2}c^{\mu_1}_{p_1}, \label{eq:casscf_eri1}
    \\
    b^P_{p_1p_2} &=& b^P_{p_1\mu_2}c^{\mu_2}_{p_2}, \label{eq:casscf_eri2}
    \\
    g^{p_1a_1}_{p_2a_2} &=& b^P_{p_1p_2} b^P_{a_1a_2}, \label{eq:casscf_eri3}
    \\
    g^{p_1p_2}_{a_1a_2} &=& b^P_{p_1a_1} b^P_{p_2a_2}, \label{eq:casscf_eri4}
\end{eqnarray}
The asymptotic operation cost scaling of evaluating these ERIs depends on the size of the active space compared to that of the AO basis: if $M_A\propto M_{\mathrm{tot}}^0$, then the first two steps, which scale the same as Eq.\ (\ref{eq:v3body}), are slowest; if $M_A\propto M_{\mathrm{tot}}^1$, the scaling is determined by the last two steps and is $O(M_{\mathrm{aux}}M_{\mathrm{tot}}^2M_A^2)$.

\paragraph{LASSCF ERIs.} The explicit ERIs required by LASSCF are those in the impurity subspace for each fragment. These are generated as
\begin{eqnarray}
    b^P_{i_{K1}\mu_2} &=& b^P_{\mu_1\mu_2}c^{\mu_1}_{i_{K1}}, \label{eq:lasscf_eri1}
    \\
    b^P_{i_{K1}i_{K2}} &=& b^P_{i_{K1}\mu_2}c^{\mu_2}_{i_{K2}}, \label{eq:lasscf_eri2}
    \\
    g^{i_{K1}i_{K3}}_{i_{K2}i_{K4}} &=& b^P_{i_{K1}i_{K2}}b^P_{i_{K3}i_{K4}}, \label{eq:lasscf_eri3}
\end{eqnarray}
which must be evaluated for all $K$. Equation (\ref{eq:lasscf_eri1}) is again isomorphic to Eq.\ (\ref{eq:v3body}) and therefore also has an operation cost scaling of $O(M_{\mathrm{aux}}M_{\mathrm{tot}}^3)$, but unlike the CASSCF ERIs, there is no possible way for the later steps to become rate limiting with a larger cost scaling, because the molecule is divided into two or more fragments and only ERIs with all fragment subindices in common are computed. LASSCF therefore drops the $O(M_{\mathrm{aux}}M_{\mathrm{tot}}^2M_A^2)$ step required by CASSCF.

\paragraph{vLASSCF ERIs.} The two additional sets of ERIs required by vLASSCF as indicated in Secs.\ \ref{sec:lasci_scf} and \ref{sec:gradhess_imp} are evaluated collectively as
\begin{eqnarray}
    b^P_{a_{K1}\mu_2} &=& b^P_{\mu_1\mu_2}c^{\mu_1}_{a_{K1}}, \label{eq:vlasscf_eri1}
    \\
    b^P_{a_{K1}a_{L1}} &=& b^P_{a_{K1}\mu_2}c^{\mu_2}_{a_{L1}}, \label{eq:vlasscf_eri2}
    \\
    g^{\mu_1a_{L1}}_{a_{K1}a_{M1}} &=& b^P_{a_{K1}\mu_1}b^P_{a_{L1}a_{M1}}, \label{eq:vlasscf_eri3}
    \\
    g^{p_1a_{L1}}_{a_{K1}a_{M1}} &=& g^{\mu_1a_{L1}}_{a_{K1}a_{M1}}c^{\mu_1}_{p_1}. \label{eq:vlasscf_eri4}
\end{eqnarray}
Here, we do explore ERIs with indices spanning multiple fragment subindices, but only when at least three of those indices are active-orbital indices. We therefore have two steps with an $O(M_{\mathrm{aux}}M_{\mathrm{tot}}^3)$ cost scaling (again, the same as the effective potentials), one step with scaling $O(M_{\mathrm{aux}}M_{\mathrm{tot}}M_A^3)$, and a final step that scales as $O(M_{\mathrm{tot}}^2M_A^3)$.

The penultimate step is potentially rate-limiting if $M_A^3>M_{\mathrm{tot}}^2$, but note that its scaling is a factor of $M_A/M_{\mathrm{tot}}$ less than that which appears in CASSCF [Eqs.\ (\ref{eq:casscf_eri3}) and (\ref{eq:casscf_eri4})]. This ratio is often less than 0.1. For instance, the organometallic test system discussed in Sec.\ \ref{sec:results_discussion} below, which has only 12 heavy atoms as ligands, has hundreds of AOs when a triple-$\zeta$ basis set is used, and no more than 14 active orbitals in MC-SCF calculations reported in the literature.\cite{Latevi2012,Wilbraham2017,Pandharkar2019} Also, as mentioned in Sec.\ \ref{sec:lasci_scf}, it is actually not strictly necessary to include ERIs coupling four independent subspaces in order to evaluate all necessary gradients, and in fact we have done so specifically in order to reduce the prefactor of the quartic-scaling cost associated with dot products involving the three-center ERIs. Therefore, future implementations may not require a quintic-scaling step at all.

The additional ERIs required by the ``full Hessian'' protocol are obtained by Eqs.\ (\ref{eq:lasscf_eri1}), (\ref{eq:lasscf_eri2}), and (using the fact that active orbitals are contained within impurity orbitals)
\begin{eqnarray}
    g^{\mu_1i_{K2}}_{i_{K1}i_{K3}} &=& b^P_{\mu_1i_{K1}}b^P_{i_{K2}i_{K3}}, \label{eq:vlasscf_eri5}
    \\
    g^{u_1i_{K2}}_{i_{K1}i_{K3}} &=& g^{\mu_1i_{K2}}_{i_{K1}i_{K3}} c^{\mu_1}_{u_1}, \label{eq:vlasscf_eri6}
\end{eqnarray} 
which have operation cost scalings of $O(n_{\mathrm{frag}}M_{\mathrm{aux}}M_{\mathrm{tot}}M_{F_K}^3)$ and $O(n_{\mathrm{frag}}M_{\mathrm{tot}}^2M_{F_K}^3)$ respectively. These scalings are at most cubic (because $M_{F_K}^3\propto M_{\mathrm{tot}}^0$) and so cannot be steeper than the quartic scaling of the effective potential evaluation.

\subsubsection{Summary of formal cost scaling analysis}

From the foregoing analysis we conclude that formally, 
\begin{enumerate}
\item if $M_A \propto M_{\mathrm{tot}}^0$, we expect 
    \begin{enumerate}
    \item CASSCF orbital optimization, LASSCF, and vLASSCF to all have asymptotically quartic $[O(M_{\mathrm{aux}}M_{\mathrm{tot}}^3)]$ operation cost scaling, and
    \item the prefactor of this scaling to increase in the order LASSCF $<$ ``approximate-Hessian'' vLASSCF $<$ ``full-Hessian'' vLASSCF, because additional steps with quartic scaling are added to the method in this order [i.e., Eqs.\ (\ref{eq:lasscf_eri1}), (\ref{eq:vlasscf_eri1}), and (\ref{eq:lasscf_eri1}) a second time], and
    \end{enumerate}
\item if $M_A \propto M_{\mathrm{tot}}^1$, we expect
    \begin{enumerate}
    \item CASSCF orbital optimization to have asymptotically quintic [$O(M_{\mathrm{aux}}M_{\mathrm{tot}}^2M_A^2)$] cost scaling,
    \item LASSCF to have the same quartic cost scaling as in the previous case, and
    \item vLASSCF to have asymptotically quintic cost scaling with a small prefactor [$O(M_{\mathrm{aux}}M_{\mathrm{tot}}M_A^3)$], such that we expect to observe the same quartic cost scaling as in the previous case in many practical uses. 
    \end{enumerate}
\end{enumerate}

\section{Results and Discussion \label{sec:results_discussion}}

All electronic structure calculations, including the impurity model CASSCF steps [Eq.\ (\ref{eq:imp_casscf})] carried out within LASSCF and vLASSCF, were performed using {\sc PySCF} version 1.7.0a.\cite{Sun2018} We implemented LASSCF and vLASSCF in the {\it mrh} software package.\cite{mrh_software} 

\begin{figure}
\includegraphics[scale=1.0]{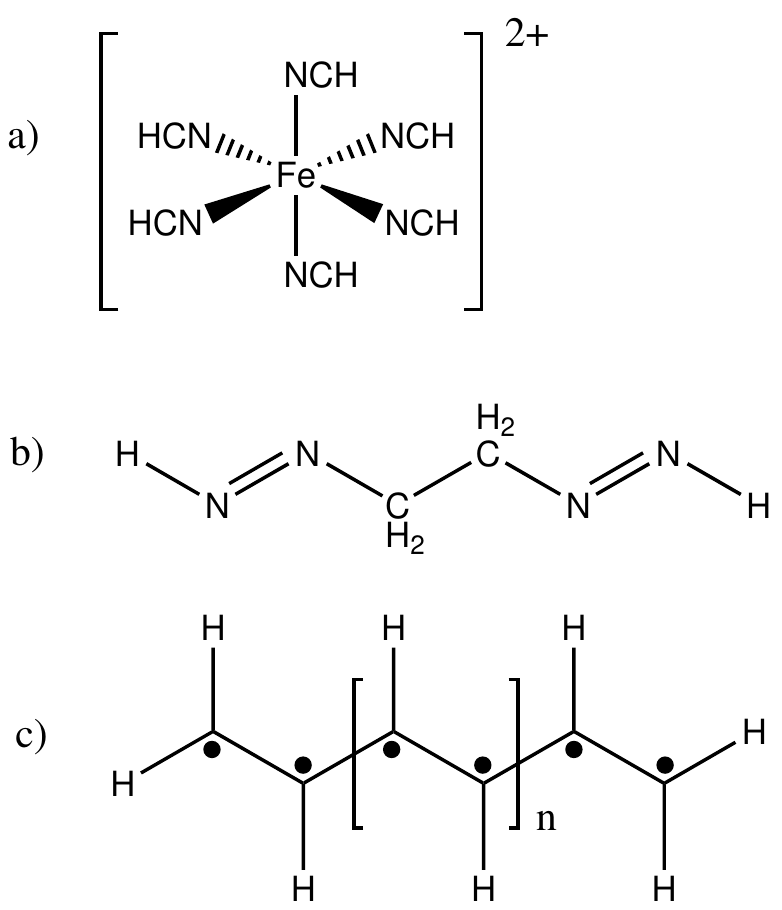}
\caption{Testbed systems for vLASSCF explored in this work: a) Model compound [Fe(NCH)$_6$]$^{2+}$, b) 2-diazenylethyldiazene (``bisdiazene''), and c) $(n+2)$-polyene chains with spin quantum numbers $s=m_s=n+2$. \label{fig:testbeds}}
\end{figure}

The testbed systems we use to explore vLASSCF are depicted in Fig.\ \ref{fig:testbeds}. The equilibrium geometries of test systems labeled a), b), and c) were obtained respectively from Refs.\ \citenum{Pandharkar2019}, \citenum{Hermes2019}, and \citenum{Hirata2004}. In the last case, the cited reference studies polyacetylene under periodic boundary conditions; we investigated this system under open boundary conditions and accordingly added terminal hydrogen atoms with a C-H bond distance of 1.091 $\AA$ and a terminal H-C-H bond angle of 123.4 degrees. For the iron complex model compound, we use the ANO-RCC-VTZP AO basis; for the other two testbed systems, we use the 6-31g AO basis. For systems a) and c), we utilize density fitting for the ERIs using an even-tempered auxiliary basis\cite{Stoychev2016} as implemented in {\sc PySCF} with the default parameter $\beta=2.0$. The purpose to study systems a) and b) was to explore how well the vLASSCF wave function can reproduce the CASSCF wave function. System c) was instead investigated to compare the timings for an analogous calculation defined both in the vLASSCF and CASSCF frameworks, as explained below. The calculations on systems a) and c), which were used to explore the methods' operation cost, were carried out on one Intel Haswell E5-2680v3 node, using 8 threads and 22 GB of RAM for system a) and 1 thread and 62 GB of RAM for system c). 

\subsection{[Fe(NCH)$_6$]$^{2+}$ and convergence to CASSCF}

The totally-symmetric singlet electronic state of the [Fe(NCH)$_6$]$^{2+}$ test system was investigated using the 3$d$ active space (6 electrons in 5 orbitals) using CASSCF, LASSCF, and various versions of vLASSCF. The electronic configuration of iron in this system is [Ar]3d$^6$ and its electronic ground state is a quintet; we enforce spin symmetry explicitly in all calculations so that no root-switching from the higher-energy singlet state can occur. In the latter two cases, the molecule was partitioned into two fragments, one containing the iron atom and all active orbitals and the other containing all ligands and no active orbitals. This partitioning of the molecule preserves its O$_{\mathrm{h}}$ point group molecular symmetry, allowing for an explicit constraint to totally symmetric wave functions in the LASSCF and vLASSCF calculations, which facilitated the numerical stability of all calculations. It also renders the LAS wave function formally equivalent to the CAS wave function.

\begin{figure}
\includegraphics[scale=1.0]{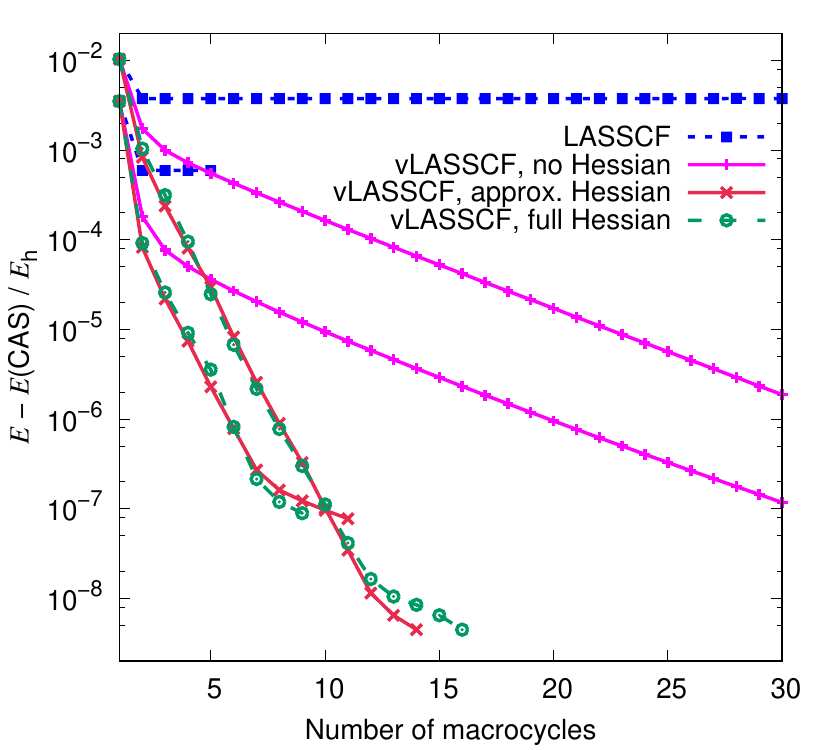}
\caption{Convergence over successive macrocycles (i.e., loops through all four decision nodes of the algorithms depicted in Figs.\ \ref{fig:oldLASSCF} and \ref{fig:vLASSCF}) of the singlet electronic energy of [Fe(NCH)$_6$]$^{2+}$ with the LASSCF(6,5)/ANO-RCC-VTZP and vLASSCF(6,5)/ANO-RCC-VTZP methods using various schemes for constructing impurity orbitals in the latter, compared to the CASSCF(6,5)/ANO-RCC-VTZP singlet energy. Two calculations of each type were carried out in which the orbitals were initialized from two different HF wave functions; the lower-energy curves of each type of calculation depict the results when initialized with RHF, and the higher-energy curves depict the results when initialized with quintet ROHF.\label{fig:fench_conv}}
\end{figure}

In Figure \ref{fig:fench_conv}, we plot the convergence of LASSCF and vLASSCF electronic energies with respect to the CASSCF(6,5) value of -1828.6865336 $E_{\mathrm{h}}$. We carry out two sets of calculations, one which uses the restricted HF (RHF) singlet to initialize the orbitals, leading to a first-cycle energy in all cases of -1828.683021 $E_{\mathrm{h}}$, and one in which the $s=m_s=2$ restricted open-shell HF (ROHF) quintet is used to initialize the orbitals, leading to a higher first-cycle energy of -1828.67622 $E_{\mathrm{h}}$ (the first-cycle energy is defined as cycle \# = 1 and corresponds to the CASCI singlet energy for the initial guess orbitals). For vLASSCF, in addition to the ``approximate Hessian'' and ``full Hessian'' protocols described in Sec.\ \ref{sec:gradhess_imp}, we explore a ``no Hessian'' protocol in which Eqs.\ (\ref{eq:hessorbs_d})--(\ref{eq:hessorbs_add}) are dropped entirely and only the gradient is used to augment the impurity space.

Even though the LAS wave function is formally equivalent to the CAS wave function, the LASSCF method is incapable of replicating the CASSCF energy due to the self-referential constraint discussed in Sec.\ \ref{sec:theory_lasscf}. LASSCF iteration does reduce the LAS energy from the first-cycle value, but it quickly plateaus and the LASSCF calculation initiated at the higher-energy point does not even succeed in reaching the lower first-cycle energy. The LASSCF iteration also fails to terminate even after 30 cycles when initiated at the higher energy, indicating that the density matrices, orbitals, or both fail to reach stable points in the constrained optimization.

On the other hand, all varieties of vLASSCF drive the LAS energy monotonically towards the CAS result, due to the lifting of all constraints on orbital relaxation. Without using the Hessian to augment the impurity, this process takes a huge number of iterations, and the overall method is untenably slow. The Hessian greatly accelerates this convergence, with vLASSCF agreeing with CASSCF's energy prediction to within 1 $\mu E_{\mathrm{h}}$ by 6 cycles for the low-spin-initiated calculations and 8 cycles for the high-spin-initiated calculations. The approximate-Hessian and full-Hessian protocols are nearly indistinguishable in terms of their effects on the number of cycles required to reach convergence. The approximate-Hessian calculations completed in slightly less wall time (24 and 35 minutes for the calculations with the low-spin and high-spin intializations respectively) than the full-Hessian calculations (25 and 41 minutes).

\subsection{Bisdiazene double-double bond dissociation potential energy curve}

The bisdiazene [system b) in Fig.\ \ref{fig:testbeds}] was partitioned into fragments and assigned active subspaces as in Ref.\ \citenum{Hermes2019}: two fragments consisting of the pairs of nitrogen bonds and their terminal hydrogens, and one fragment consisting of the central C$_2$H$_4$ unit. Active spaces of (4,4) were assigned to the terminal diazene fragments and no active space was assigned to the central unit, for an overall CAS of (8,8). The potential energy surface was scanned along the simultaneous stretching coordinate of the two N=N double bonds from the reference equilibrium geometry, which is reported in the SI of Ref.\ \citenum{Hermes2019}.

\begin{figure}
\includegraphics[scale=1.0]{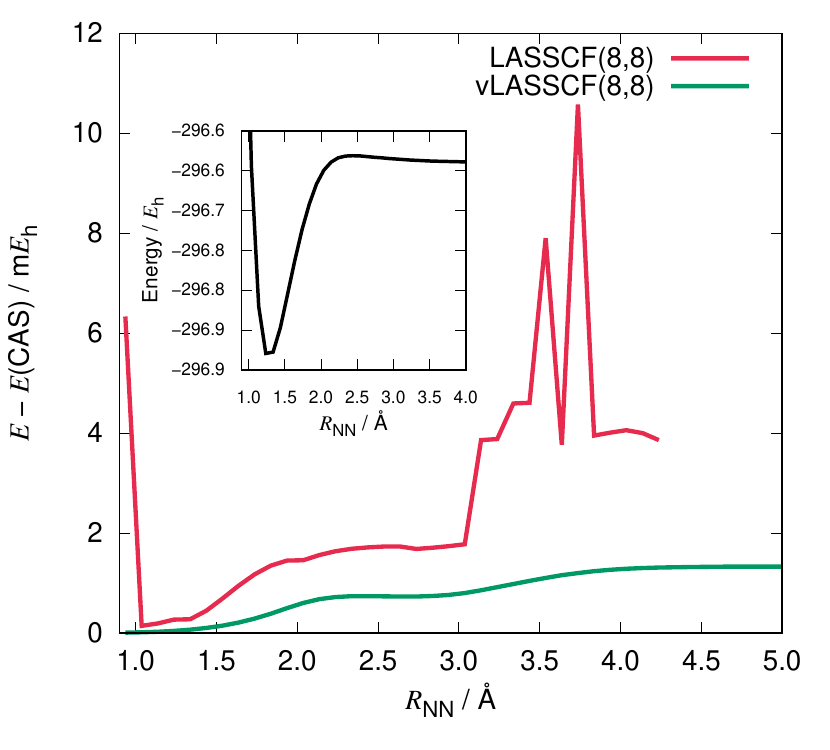}
\caption{Energy difference between LASSCF(8,8), vLASSCF(8,8), CASSCF(8,8) models of bisdiazene in the 6-31g basis across the simultaneous nitrogen-nitrogen double-bond dissociation coordinate. The LASSCF data is taken from Ref.\ \citenum{Hermes2019}. Inset: the CASSCF(8,8)/6-31g potential energy curve. \label{fig:c2h6n4_pes}}
\end{figure}

Figure \ref{fig:c2h6n4_pes} shows the difference between the LASSCF and vLASSCF energies and the CASSCF reference along this potential energy curve. (The vLASSCF result is the same regardless of whether or how the Hessian was used to augment the impurity.) Because there are two active subspaces with non-trivial wave functions, LAS and CAS wave functions are not formally equivalent. Nevertheless, because the two active subspaces are physically separated and not entangled (e.g., \emph{via} a connecting $\pi$-orbital system), the LAS wave function is only a mild approximation to the CAS wave function. For this reason, both LASSCF and vLASSCF predict total electronic energies within a few m$E_{\mathrm{h}}$ throughout most of the potential energy surface.

However, the LASSCF potential energy curve is significantly less smooth (as well as everywhere higher) than the vLASSCF potential energy curve.  Near the equilibrium geometry, the curve is reasonably smooth, but past $R_{\mathrm{N}=\mathrm{N}}=3.0\ \AA$, it becomes highly erratic and discontinuous, despite the fact that orbitals from converged LASSCF calculations at one point were used to initialize calculations at another point only 0.1 $\AA$ further out or in. The discontinuities in the LASSCF curve demonstrate the consequences of the use of ill-defined, self-referential constraints in variational optimizations.

\begin{figure}
\includegraphics[scale=1.0]{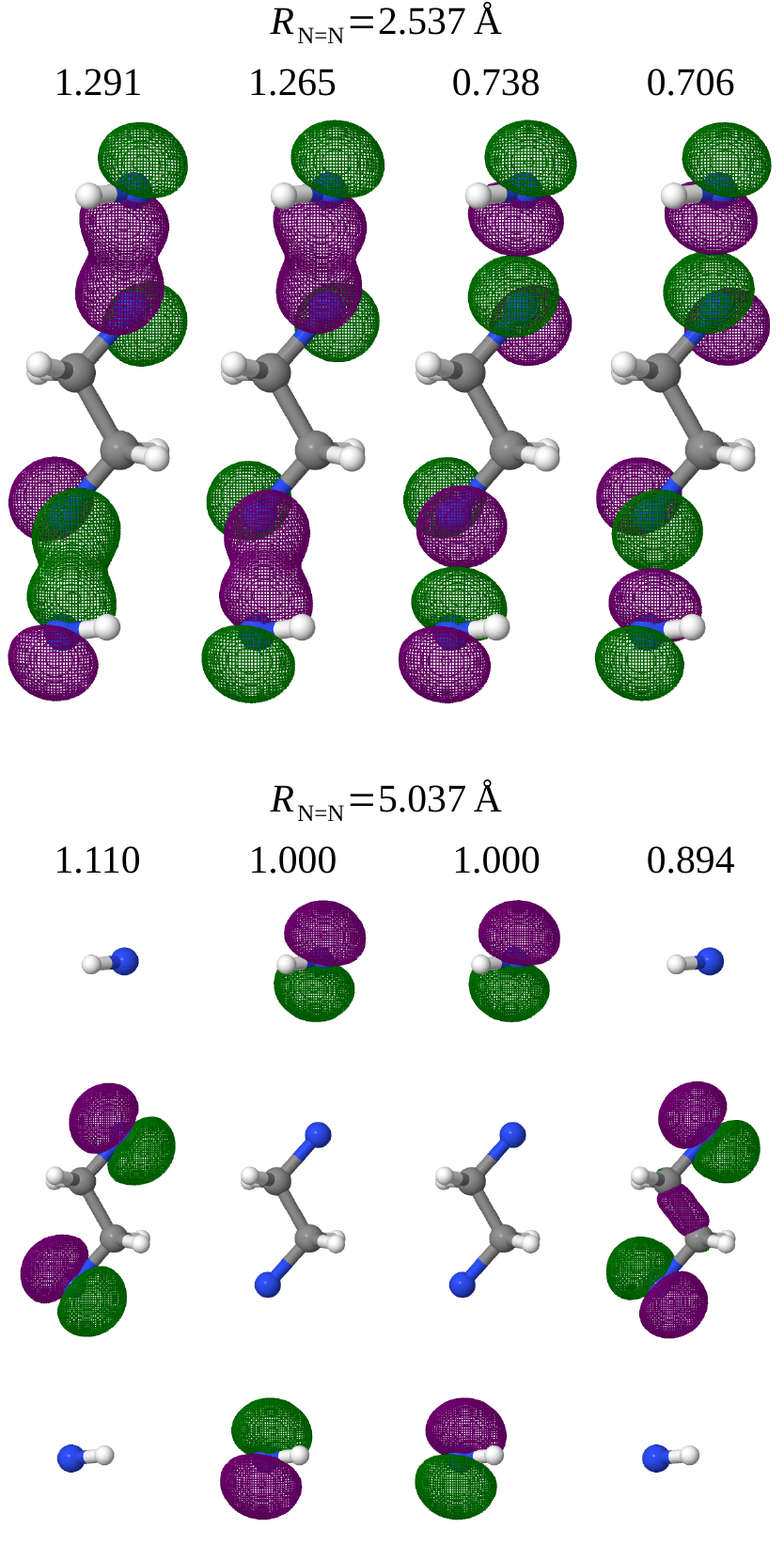}
\caption{Selected CASSCF(8,8)/6-31g natural orbitals and their occupancies for bisdiazene at two geometries along the simultaneous nitrogen-nitrogen double-bond dissociation coordinate. \label{fig:c2h6n4_nos}}
\end{figure}

The vLASSCF extension entirely cures these defects of the LASSCF potential energy curve. Three regions are visible, corresponding to 1) the neighborhood of the equilibrium geometry, where the molecule is weakly correlated and vLASSCF and CASSCF are nearly indistinguishable, 2) a plateau at about $E_{\mathrm{LAS}}-E_{\mathrm{CAS}}=0.7$ m$E_\mathrm{h}$ near the potential energy maximum along the dissociation coordinate, and 3) a second plateau of about 1.3 m$E_{\mathrm{h}}$ in the dissociation limit (confirmed out to $R_{\mathrm{N}=\mathrm{N}}\approx 100\ \AA$). The transition between the first plateau and the second corresponds to the transition in the character of the CASSCF natural orbitals between pairs of bonding and anti-bonding $\sigma$ and $\pi$ orbitals at shorter distances, and isolated nitrogen $p$ orbitals with more entanglement across the central fragment at longer distances, as depicted in Fig.\ \ref{fig:c2h6n4_nos}. In the first plateau, around $R_{\mathrm{N}=\mathrm{N}}=2.5\ \AA$, the natural orbitals consist of four pairs of bonding and anti-bonding orbitals with NO occupancies differing by 0.03 to 0.04 between the in-phase combination (first and third orbitals in the top row of Fig.\ \ref{fig:c2h6n4_nos}) and the out-of-phase combination (second and fourth). On the other hand, in the dissociation limit, one pair of CASSCF natural orbitals consist of the in-phase and out-of phase combinations of two nitrogen atom $p$ orbitals in the two different fragments, with a difference in natural occupancies of more than 0.2. This greater difference in natural-orbital occupancies implies greater entanglement, and consequently that localization and approximation as unentangled fragments is a slightly more severe approximation in the latter case than in the former, corresponding to an additional variational penalty of about 0.6 m$E_{\mathrm{h}}$.

\subsection{Wall time required for high-spin polyene chain calculations}

\begin{figure*}[ht]
\includegraphics[scale=1.0]{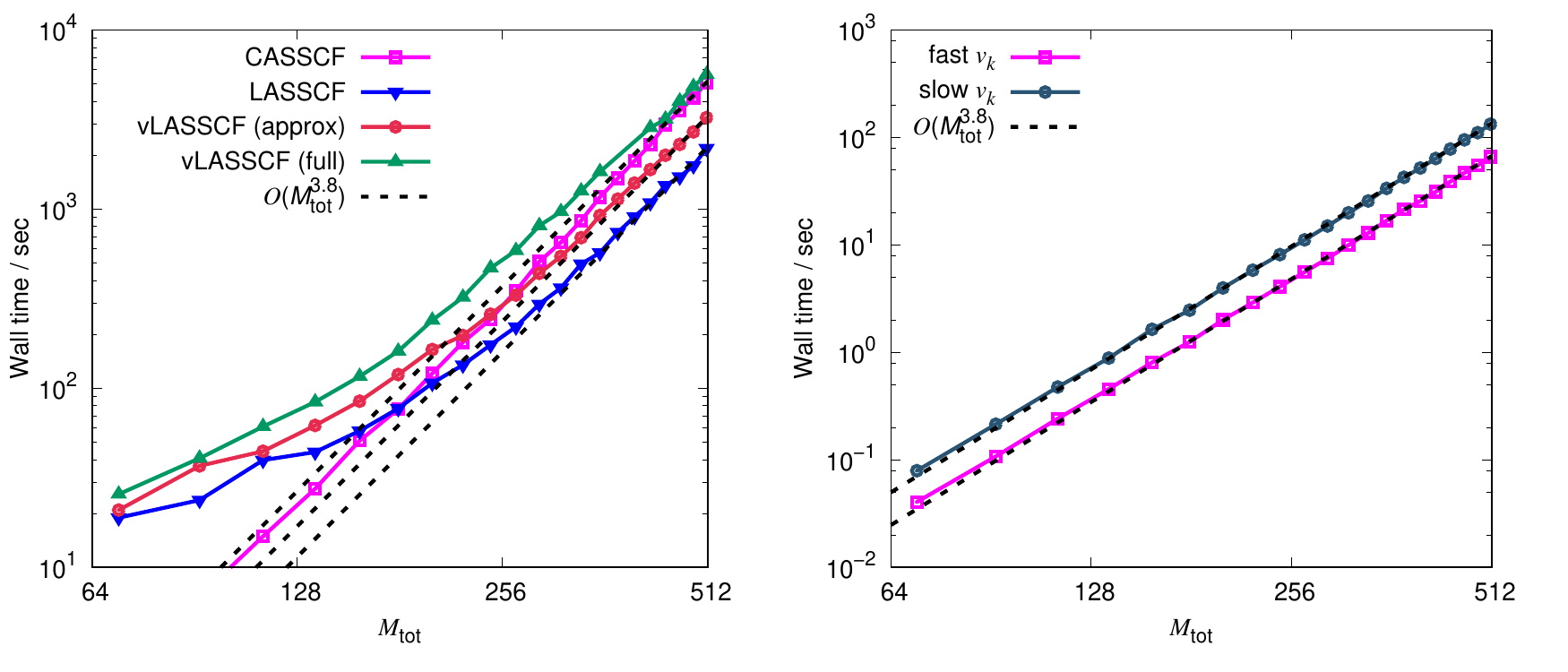}
\caption{Left: Wall time required to complete CASSCF, LASSCF, and vLASSCF calculations of high-spin states of $(n+2)$-polyene chains with $1\le n<22$ in the 6-31g basis plotted against the number of atomic orbitals. Right: Average wall time required to calculate the Coulomb and exchange effective potential matrices using the two available standard implementations in {\sc PySCF}. Analytical functions $a\times M_{\mathrm{tot}}^{3.8}$ with various constant factors $a$ are also plotted in both figures as guides to the eye.
\label{fig:polyene_profiling}}
\end{figure*}

The operation costs of LASSCF and vLASSCF compared to CASSCF were examined numerically using a model system [system c) in Fig.\ \ref{fig:testbeds}] in which the $\pi$ electrons of finite all-\emph{trans} polyenes of increasing length were all assigned the same spin ($n+2=s=m_s$), and a complete active space of 1 orbital and 1 electron per carbon atom was used and divided into fragments consisting of two CH units each in the LASSCF and vLASSCF calculations. This highly artificial system has only one determinant in its complete active space, meaning that, formally, the LAS, CAS, and ROHF wave function forms are all equivalent, and the CASSCF CI vector is only one determinant long no matter how many active orbitals there are. This allows us to compare the computational costs of LASSCF and vLASSCF to the cost of the \emph{orbital optimization part} of a CASSCF calculation directly. Table S10 of the SI presents the total electronic energies of the calculations reported in Fig.\ \ref{fig:polyene_profiling} and confirms the energetic equivalence of CASSCF and vLASSCF in this case.

Figure \ref{fig:polyene_profiling} plots the wall time required by CASSCF, LASSCF, and vLASSCF calculations (using both full-Hessian and approximate-Hessian protocols) on the high-spin all-\emph{trans} polyenes of increasing length, compared to the number of atomic orbitals ($M_{\mathrm{tot}}$) of the system, on logarithmic axes. All varieties of LASSCF (in their current implementations) are slower than CASSCF for small systems due to low-scaling overhead steps. However, the cost of LASSCF relative to the cost of CASSCF orbital optimization decreases with increasing size, and by $M_{\mathrm{tot}}=250$, both LASSCF and approximate-Hessian vLASSCF are faster than CASSCF.

Recall from Sec.\ \ref{sec:theory_cost_formal} that all methods here explored include the repeated calculation of effective potential matrices with formal operation cost scaling of $O(M_{\mathrm{aux}}M_{\mathrm{tot}}^3)$. The scaling of this step is explored numerically in the right-hand panel of Fig.\ \ref{fig:polyene_profiling}, and appears to be approximately $O(M_{\mathrm{tot}}^{3.8})$ in the size range explored here. The measured operation cost scalings of CASSCF and all forms of LASSCF appear to be close to this value in the region of $M_{\mathrm{tot}}=500$, but that of CASSCF appears slightly higher, and those of LASSCF and vLASSCF appear slightly lower. In Sec. \ref{sec:theory_cost_formal}, we concluded that vLASSCF and CASSCF should both have a formally quintic asymptotic cost scaling in the case that $M_A\propto M_{\mathrm{tot}}^1$ (which is the case here), but that the quintic step in vLASSCF is faster than the quintic step in CASSCF by a factor of $M_A/M_{\mathrm{tot}}$. Here, that ratio is $1/11\approx 0.09$, which is small enough to suppress any evidence of quintic cost scaling for up to 500 orbitals. The prefactor associated with the measured quartic cost scaling increases in the order LASSCF $<$ approximate-Hessian vLASSCF $<$ full-Hessian vLASSCF, consistent with the analysis in Sec.\ \ref{sec:theory_cost_formal}.

Although the improvement in computational cost of vLASSCF as currently implemented is modest, note that it is dominated by the calculation of a Hartree--Fock like exchange potential, meaning that any improvement in the scaling of Hartree--Fock calculations (such as, for instance, a linear-scaling implementation of exchange potential calculation\cite{Koppel2016}) can be applied to LASSCF and vLASSCF as well for immediate further speedup. So long as the formal quintic-scaling step in vLASSCF remains insignificant (or is removed in a future implementation as alluded to at the end of Sec.\ \ref{sec:lasci_scf}), vLASSCF does not include any step with worse operation cost scaling than Hartree--Fock.

\section{Conclusions}

Our recently-developed LASSCF method defeats the exponential cost scaling of CASSCF with respect to the size of the active space by splitting the active space into unentangled fragments and uses a DMET-inspired algorithm to additionally break the orbital optimization process into many short steps. However, in the original theory and implementation, an ill-defined system of constraints on the optimization of the active orbitals limits its robustness and reproducibility. The variational version of the method introduced here, which we have named vLASSCF, cures these deficiencies and is truly variational in the Hellmann-Feynman sense, which improves upon the consistency and transferability of the method and allows for more trustworthy analysis of small energy differences. We have therefore jettisoned some of the baggage of DMET (i.e., dependence on user choice of orbital localization protocol) while retaining the attractive feature of splitting an MC-SCF orbital optimization problem into several small coupled optimization problems. The superior smoothness of the bisdiazene potential energy curve, the confirmation of energetic equivalence between CASSCF and vLASSCF in the appropriate limit, and our formal and analytical operation cost analyses collectively demonstrate that we have succeeded in having our cake and eating it too.

In addition to the improved quality of the vLASSCF results compared to LASSCF, vLASSCF is amenable to the straightforward calculation of molecular gradients using the Hellmann-Feynman theorem.  We have already shown that LASSCF is an attractive alternative to CASSCF in the calculation of spin-state energetics,\cite{Pandharkar2019} in which the separable form of LASSCF facilitates chemical interpretation of the wave function in a manner that is often obscured by the CAS formalism. The variational formalism and forthcoming gradient implementation will allow LASSCF to explore the relationship between spin state energetics and molecular geometries as well. Finally, LAS wave functions of more robust accuracy offered by vLASSCF are expected to be critically important in the context of post-SCF methods such as MC-PDFT\cite{Manni2014,Carlson2015,Gagliardi2017,Wilbraham2017,Sand2018} as well as post-LAS wave function formalisms such as ASD.\cite{Parker2013,Parker2014,Parker2014a,Parker2014b,Kim2015} It will be interesting in the future to test the method on systems containing multiple metal and/or lanthanide/actinide centers.

\acknowledgement
This work was in part funded by the Division of Chemical Sciences, Geosciences, and
Biosciences, Office of Basic Energy Sciences of the U.S. Department of Energy through grant USDOE/DESC002183.

\begin{suppinfo}
Additional programmable equations, absolute electronic energies, calculation wall times, and equilibrium molecular geometry Cartesian coordinates.
\end{suppinfo}

\begin{tocentry}
\includegraphics{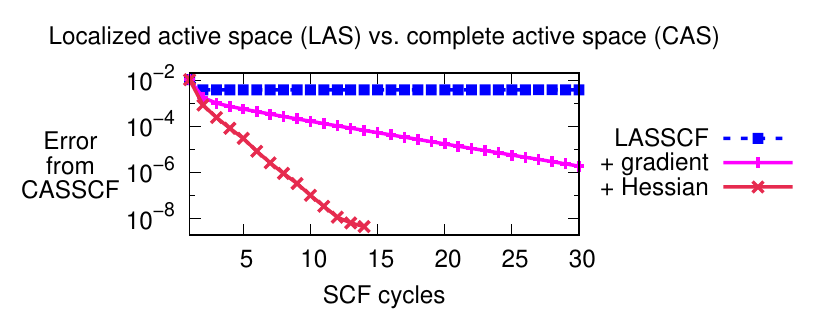}
\end{tocentry}

\bibliography{draft3.bib}
\end{document}


\section{Gradient and Hessian expressions}

\subsection{Formal conventions\label{sec:conventions}}

In this document, the orbital indices $p_n$ with integer subindices $n=1,2,3,\ldots$ range over an arbitrary set of orthonormal MOs. Orbital indices $q_n, r_n, s_n,$ and $t_n$ identify any unentangled orbital subspace of the molecule [for example, the inactive orbital ($\ket{d_n}$), virtual orbital ($\ket{v_n}$), unactive (inactive + virtual) orbital ($\ket{u_n}$), or active/impurity orbital sets associated with any fragment ($\ket{a_{Kn}}/\ket{i_{Kn}}$, $\ket{a_{Ln}}/\ket{i_{Ln}}$, etc.)].

Evaluating energy derivatives with respect to wave function transformations requires parameterizing a unitary operator, 
\begin{eqnarray}
    E_{\mathrm{LAS}} &=& \braket{\mathrm{LAS}|\hat{U}_{\mathrm{LAS}}^{\dagger} \hat{H} \hat{U}_{\mathrm{LAS}}|\mathrm{LAS}}, \label{eval_umat}
\end{eqnarray}
in terms of weighted sums of generators and differentiating the expectation value expression with respect to the amplitudes of the generators. For a LAS wave function, the unitary operator is separable into a product of operators for the orbitals and the CI vectors of the active subspaces,
\begin{eqnarray}
    \hat{U}_{\mathrm{LAS}} &=& \hat{U}_{\mathrm{orb}}\prod_K \hat{U}_{\mathrm{CI},K},
\end{eqnarray}
so long as redundant orbital rotations are omitted from $\hat{U}_{\mathrm{orb}}$. Unitary operators can be constructed as exponentials of weighted sums of anti-hermitian generators,
\begin{eqnarray}
    \hat{U}_{\mathrm{orb}} &=& \mathrm{exp}\left[\sum_{q>r}\sum_{q_1,r_1} \vecX^{q_1}_{r_1}\left(\crop{q_1\sigma}\anop{r_1\sigma}-\crop{r_1\sigma}\anop{q_1\sigma}\right)\right],
    \\
    \hat{U}_{\mathrm{CI},K} &=& \mathrm{exp}\left[\sum_{\vec{a}_{K1}} \vecX_{\vec{a}_{K1}} \left(\ket{\vec{a}_{K1}}\bra{\Psi_{A_K}}-\ket{\Psi_{A_K}}\bra{\vec{a}_{K1}}\right)\right].
\end{eqnarray}

\subsection{Explicit gradient and Hessian expressions for $\ket{\mathrm{LAS}}$}

The first and second derivatives of Eq.\ (\ref{eval_umat}) with respect to the generator amplitudes $\vecX^{q_1}_{r_1}$ and $\vecX_{\vec{a}_{K1}}$ are
\begin{eqnarray}
    \Emat{1}^{q_1}_{r_1} &=& \Fmat{1}^{q_1}_{r_1} - \Fmat{1}^{r_1}_{q_1}, \label{eq:Emat1}
    \\
    \Emat{2}^{q_1s_1}_{r_1t_1} &=& \Fmat{2}^{q_1s_1}_{r_1t_1} - \Fmat{2}^{r_1s_1}_{q_1t_1} - \Fmat{2}^{q_1t_1}_{r_1s_1} + \Fmat{2}^{r_1t_1}_{q_1s_1}
    \nonumber \\ && + \frac{1}{2}\left(\braket{r_1|s_1}\left[\Fmat{1}^{q_1}_{t_1}+\Fmat{1}^{t_1}_{q_1}\right]
    - \braket{r_1|t_1}\left[\Fmat{1}^{q_1}_{s_1}+\Fmat{1}^{s_1}_{q_1}\right]\right.
    \nonumber \\ &&           \left. - \braket{q_1|s_1}\left[\Fmat{1}^{r_1}_{t_1}+\Fmat{1}^{t_1}_{r_1}\right]
    + \braket{q_1|t_1}\left[\Fmat{1}^{r_1}_{s_1}+\Fmat{1}^{s_1}_{r_1}\right]\right), \label{eq:Emat2}
    \\
    \Evec{1}_{\vec{a}_{K1}} &=& \braket{\vec{a}_{K1}|\hat{Q}_{A_K}\hat{H}_{A_K}|\Psi_{A_K}} + \mathrm{h.c.}, \label{eq:Evec1}
    \\
    \Evec{2}_{\vec{a}_{K1}\vec{a}_{K2}} &=& \braket{\vec{a}_{K1}|\hat{Q}_{A_K}\left(\hat{H}_{A_K}-E_{A_K}\right)\hat{Q}_{A_K}|\vec{a}_{K2}} + \mathrm{h.c.}, \label{eq:Evec2d}
    \\
    \Evec{2}_{\vec{a}_{K1}\vec{a}_{L1}} &=& 
    \braket{\vec{a}_{K1}|\hat{Q}_{A_K}\hat{H}_{A_K}^{(\vec{a}_{L1})}|\Psi_{A_K}} + \mathrm{h.c.}
    =\braket{\vec{a}_{L1}|\hat{Q}_{A_L}\hat{H}_{A_L}^{(\vec{a}_{K1})}|\Psi_{A_L}} + \mathrm{h.c.}, \label{eq:Evec2o}
    \\
    \{\Evec{2}_{\vec{a}_{K1}}\}^{q_1}_{r_1} &=& \{\Fvec{2}_{\vec{a}_{K1}}\}^{q_1}_{r_1} - \{\Fvec{2}_{\vec{a}_{K1}}\}^{r_1}_{q_1}, \label{eq:Evec2c}
\end{eqnarray}
where $\braket{q_1|r_1}$ is an orbital overlap integral; $\hat{H}_{A_K}$, $\hat{Q}_{A_K}$, and ``h.c.'' are defined in the main text; and
\begin{eqnarray}
    \Fmat{1}^{q_1}_{r_1} &=& h^{q_1}_{p_1} D^{r_1}_{p_1} + g^{q_1p_2}_{p_1p_3} d^{r_1p_2}_{p_1p_3} \label{eq:Fmat1}
    \\
    \Fmat{2}^{q_1s_1}_{r_1t_1} &=& h^{q_1}_{s_1} D^{r_1}_{t_1} + g^{q_1s_1}_{p_1p_2} d^{r_1t_1}_{p_1p_2} + g^{q_1p_1}_{s_1p_2} d^{r_1p_1}_{t_1p_2} + g^{q_1p_1}_{p_2s_1} d^{r_1p_1}_{p_2t_1} \label{eq:Fmat2}
    \\
    E_{A_K} &=& \braket{\Psi_{A_K}|\hat{H}_{A_K}|\Psi_{A_K}},
    \label{eq:Eak}
    \\
    \hat{H}_{A_K}^{(\vec{a}_{L1})} &=& \OBMal{v_\sigma}^{a_{K1}}_{a_{K2}}\crop{a_{K1}}\anop{a_{K2}}, \label{eq:Hakal}
    \\
    \{\Fvec{2}_{\vec{a}_{K1}}\}^{q_1}_{r_1} &=& \braket{r_1|a_{K1}}\left(h^{q_1}_{a_{K2}}\OBMak{D}^{a_{K1}}_{a_{K2}} + g^{q_1p_2}_{p_1p_3} \OBMak{d}^{a_{K1}p_2}_{p_1p_3}\right), \label{eq:Fvecak}
\end{eqnarray}
where $h$, $g$, $D$, and $\gamma_\sigma$ are defined in the main text, $d$ is the 2-RDM, and 
\begin{eqnarray}
    \OBMak{v_\sigma}^{p_1}_{p_2} &=& g^{p_1a_{K1}}_{p_2a_{K2}}\OBMak{D}^{a_{K1}}_{a_{K2}} 
    -g^{p_1a_{K1}}_{a_{K2}p_2}\OBMak{\gamma_\sigma}^{a_{K1}}_{a_{K2}}, \label{eq:vak}
    \\
    \OBMak{\gamma_\sigma}^{a_{K1}}_{a_{K2}} &\equiv& \left(\braket{\vec{a}_{K1}|\hat{Q}_{A_K}\crop{a_{K1}\sigma}\anop{a_{K2}\sigma}|\Psi_{A_K}} + \mathrm{h.c.}\right), \label{eq:1tdmak}
    \\
    \OBMak{d}^{a_{K1}p_2}_{p_1p_3} &\equiv& \left(\braket{\vec{a}_{K1}|\hat{Q}_{A_K}\crop{a_{K1}\sigma}\crop{p_2\tau}\anop{p_3\tau}\anop{p_1\sigma}|\Psi_{A_K}} + \mathrm{h.c.}\right), \label{eq:2tdmak}
\end{eqnarray}
where it should be understood that in both Eq.\ (\ref{eq:1tdmak}) and Eq.\ (\ref{eq:2tdmak}), at least one bra index (i.e., creation operator, $\hat{c}^\dagger$) and one ket index (i.e., annihilation operator, $\hat{c}$) must identify an active orbital from the $K$th subspace ($\ket{a_{Kn}}$). In Eq.\ (\ref{eq:2tdmak}), for any orbital indices outside of the $K$th active subspace, the expectation value in $\ket{\mathrm{LAS}}$ is taken. Equations (\ref{eq:1tdmak}) and (\ref{eq:2tdmak}) describe transition density matrices.

The cumulant decomposition,
\begin{eqnarray}
    \lambda^{p_1p_3}_{p_2p_4} &\equiv& d^{p_1p_3}_{p_2p_4} - D^{p_1}_{p_2} D^{p_3}_{p_4} + \{\gamma_\sigma\}^{p_1}_{p_4}\{\gamma_\sigma\}^{p_3}_{p_2}, \label{eq:cumulant}
\end{eqnarray}
can be generalized to the transition density matrix,
\begin{eqnarray}
    \OBMak{\lambda}^{p_1p_3}_{p_2p_4} &\equiv& \OBMak{d}^{p_1p_3}_{p_2p_4} - \OBMak{D}^{p_1}_{p_2} D^{p_3}_{p_4} - D^{p_1}_{p_2} \OBMak{D}^{p_3}_{p_4} 
    \nonumber \\ && + \OBMak{\gamma_\sigma}^{p_1}_{p_4}\{\gamma_\sigma\}^{p_3}_{p_2} + \{\gamma_\sigma\}^{p_1}_{p_4}\OBMak{\gamma_\sigma}^{p_3}_{p_2}, \label{eq:tcumulant}
\end{eqnarray}
in which terms Eqs.\ (\ref{eq:Fmat1}), (\ref{eq:Fmat2}), and (\ref{eq:Fvecak}) can be rewritten:
\begin{eqnarray}
    \Fmat{1}^{q_1}_{r_1} &=& h^{q_1}_{p_1} D^{r_1}_{p_1} + \{v_\sigma^{(jk)}\}^{q_1}_{p_1} \{\gamma_\sigma\}^{r_1}_{p_1} + g^{q_1p_2}_{p_1p_3} \lambda^{r_1p_2}_{p_1p_3}, \label{eq:Fmat1_cg}
    \\
    \Fmat{2}^{q_1s_1}_{r_1t_1} &=& h^{q_1}_{s_1} D^{r_1}_{t_1} + \{v_\sigma^{(jk)}\}^{q_1}_{s_1} \{\gamma_\sigma\}^{r_1}_{t_1} + g^{q_1s_1}_{p_1p_2} \lambda^{r_1t_1}_{p_1p_2} + g^{q_1p_1}_{s_1p_2} \lambda^{r_1p_1}_{t_1p_2} + g^{q_1p_1}_{p_2s_1} \lambda^{r_1p_1}_{p_2t_1}
    \nonumber \\ && + 2g^{q_1s_1}_{p_1p_2}D^{r_1}_{p_1}D^{t_1}_{p_2} - \left(g^{q_1s_1}_{p_2p_1} + g^{q_1p_2}_{s_1p_1}\right)\{\gamma_\sigma\}^{r_1}_{p_1}\{\gamma_\sigma\}^{t_1}_{p_2}, \label{eq:Fmat2_cg}
    \\
    \{\Fvec{2}_{\vec{a}_{K1}}\}^{q_1}_{r_1} &=& h^{q_1}_{p_1}\OBMak{D}^{r_1}_{p_1}
    + \{v_\sigma^{(jk)}\}^{q_1}_{p_1}\OBMak{\gamma_\sigma}^{r_1}_{p_1}
    + \OBMak{v_\sigma}^{q_1}_{p_1}\{\gamma_\sigma\}^{r_1}_{p_1}
    + g^{q_1p_2}_{p_1p_3}\OBMak{\lambda}^{r_1p_2}_{p_1p_3}. \label{eq:Fvecak_cg}
\end{eqnarray}

\subsection{Hessian-vector products}

The dot product of the Hessian with the unitary group generator amplitude vector $\vecX$ can be expressed compactly as 
\begin{eqnarray}
    \EmatX{2}^{q_1}_{r_1} &=& \FmatX{1}^{q_1}_{r_1} - \FmatX{1}^{r_1}_{q_1}, \label{eq:Hmatvec_aorb}
    \\
    \EmatX{2}_{\vec{a}_{K1}} &=& 2\braket{\vec{a}_{K1}|\hat{Q}_{\Psi_{A_K}}\left(\hat{H}_{A_K}-E_{A_K}\right)\hat{Q}_{\Psi_{A_K}}|x_K}
    + \braket{\vec{a}_{K1}|\hat{Q}_{\Psi_{A_K}}\hat{H}_{A_K}^{(\vecX)}|\Psi_{A_K}}, \label{eq:Hmatvec_CI}
\end{eqnarray}
with
\begin{eqnarray}
    \FmatX{1}^{q_1}_{r_1} &=& \left(h^{q_1}_{p_1} + \{v_\sigma^{(jk)}\}^{q_1}_{p_1}\right) \{\tilde{\gamma}_{\sigma}\}^{p_1}_{r_1}
    + \{\tilde{v}_\sigma^{(jk)}\}^{q_1}_{r_2} \{\gamma_{\sigma}\}^{r_2}_{r_1}
    + g^{q_1p_2}_{p_1p_3} \tilde{\lambda}^{r_1p_2}_{p_1p_3}, \label{eq:FmatX}
    \\
    \hat{H}_{A_K}^{(\vecX)} &=& \left(\tilde{h}^{a_{K1}}_{a_{K2}} + \{\tilde{v}_{\sigma}^{(jk)}\}^{a_{K1}}_{a_{K2}} - \{\tilde{v}_{\sigma}^{(\mathrm{self})}\}^{a_{K1}}_{a_{K2}}\right)\crop{a_{K1}\sigma}\anop{a_{K2}\sigma}
    +\frac{1}{2}
    \tilde{g}^{a_{K1}a_{K3}}_{a_{K2}a_{K4}}
    \crop{a_{K1}\sigma}\crop{a_{K3}\tau}\anop{a_{K4}\tau}\anop{a_{K2}\sigma}, \label{eq:KAK}
    \\
    \ket{x_K} &\equiv& \sum_{\vec{a}_{K1}} \ket{\vec{a}_{K1}}\vecX_{\vec{a}_{K1}},
\end{eqnarray}
where the transformed transformed 1-RDMs ($\tilde{D}=\tilde{\gamma}_\uparrow+\tilde{\gamma}_\downarrow$), 2-RDM cumulants ($\tilde{\lambda}$), Hamiltonian matrix elements ($\tilde{h}$ and $\tilde{g}$) and effective potentials ($\tilde{v}$) are
\begin{eqnarray}
    \{\tilde{\gamma}_\sigma\}^{p_1}_{p_2} &=& \vecX^{p_1}_{p_3} \{\gamma_\sigma\}^{p_3}_{p_2} - \{\gamma_\sigma\}^{p_1}_{p_3} \vecX^{p_3}_{p_2} + \OBMx{\gamma_\sigma}^{p_1}_{p_2}, \label{eq:tildegamma_full}
    \\
    \tilde{\lambda}^{p_1p_3}_{p_2p_4} &=& \vecX^{p_1}_{p_5} \lambda^{p_5p_3}_{p_2p_4} - \lambda^{p_1p_3}_{p_5p_4} \vecX^{p_5}_{p_2}
    + \vecX^{p_3}_{p_5} \lambda^{p_1p_5}_{p_2p_4} - \lambda^{p_1p_3}_{p_2p_5} \vecX^{p_5}_{p_4} + \OBMx{\lambda}^{p_1p_3}_{p_2p_4}, \label{eq:tildelambda_full}
    \\
    \tilde{h}^{p_1}_{p_2} &=& h^{p_1}_{p_3} \vecX^{p_3}_{p_2} - \vecX^{p_1}_{p_3} h^{p_3}_{p_2}, \label{eq:tildeh}
    \\
    \tilde{g}^{p_1p_3}_{p_2p_4} &=& g^{p_1p_3}_{p_2p_5} \vecX^{p_5}_{p_4} - \vecX^{p_1}_{p_5} g^{p_5p_3}_{p_2p_4}
    + g^{p_1p_3}_{p_5p_4} \vecX^{p_5}_{p_2} - \vecX^{p_3}_{p_5} g^{p_1p_5}_{p_2p_4}, \label{eq:tildeg}
    \\
    \{\tilde{v}_\sigma^{(jk)}\}^{\mu_1}_{\mu_2} &=& g^{\mu_1\mu_3}_{\mu_2\mu_4}\tilde{D}^{\mu_3}_{\mu_4} - g^{\mu_1\mu_3}_{\mu_4\mu_2}\{\tilde{\gamma}_\sigma\}^{\mu_3}_{\mu_4}, \label{eq:tildevjk_full}
    \\
    \{\tilde{v}_\sigma^{(\mathrm{self})}\}^{a_{K1}}_{a_{K2}} &=& 
    \sum_{L,M}\left(g^{a_{K1}a_{L1}}_{a_{K2}a_{M1}}\tilde{D}^{a_{L1}}_{a_{M1}} - g^{a_{K1}a_{L1}}_{a_{M1}a_{K2}}\{\tilde{\gamma}_\sigma\}^{a_{L1}}_{a_{M1}}\right) 
    - \sum_{L\neq K}\left(\tilde{g}^{a_{K1}a_{L1}}_{a_{K2}a_{L2}}D^{a_{L1}}_{a_{L2}} - \tilde{g}^{a_{K1}a_{L1}}_{a_{L2}a_{K2}}\{\gamma_\sigma\}^{a_{L1}}_{a_{L2}}\right) 
    \nonumber \\ &&
    - 2\sum_{L\neq K}\left(g^{a_{K1}a_{L1}}_{a_{K2}a_{L2}}\OBMxl{D}^{a_{L1}}_{a_{L2}} - g^{a_{K1}a_{L1}}_{a_{L2}a_{K2}}\OBMxl{\gamma_\sigma}^{a_{L1}}_{a_{L2}}\right), \label{eq:tildevself}
\end{eqnarray}
and where transition density matrices $\OBMxnot{D}=\OBMxnot{\gamma_\uparrow}+\OBMxnot{\gamma_\downarrow}$ and $\OBMxnot{\lambda}$ are given by Eqs.\ (\ref{eq:1tdmak}), (\ref{eq:2tdmak}), and (\ref{eq:tcumulant}) with the substitution $\ket{\vec{a}_{K1}} \to \ket{x_K}$. These expressions are valid for the entire Hessian-vector product with an arbitrary shift vector $\vecX$; however, in LASSCF, we do not minimize the energy with respect to all variables simultaneously. These equations are therefore amenable to further simplification which is left as an exercise for the reader.

\subsection{Analysis of required ERIs\label{sec:eri_analysis}}

Both the 1-RDM ($D=\gamma_\uparrow+\gamma_\downarrow$) and the cumulant ($\lambda$), as well as their transition analogues ($D^{(\vec{a}_{K1})}$, etc.) are block-diagonal in unentangled subspaces of the molecule. An unentangled subspace is a Fock space of orbitals (say $\ket{q_n}$) for which the wave function has a factorizable part:
\begin{eqnarray}
    \ket{\Psi} &=& \ket{\Psi(q_n)}\otimes\ket{\Psi(q_n^\prime)}, \label{eq:factor}
\end{eqnarray}
where $\ket{\Psi(q_n)}$ denotes a wave function defined in the Fock space of the orbital set $\{\ket{q_n}\}$, of which $\{\ket{q_n^\prime}\}$ is the orthogonal complement. (We consider wave functions that observe particle number symmetry, $\hat{N}\ket{\Psi}=N\ket{\Psi}$.) The 1-RDMs have no nonzero elements coupling two non-overlapping unentangled subspaces because that would imply that the wave function contains terms characterized by single-electron excitations like $\ket{q_1}\to\ket{q_1^\prime}$, which is incompatible with Eq.\ (\ref{eq:factor}). Likewise, the cumulant and transition cumulant have no nonzero elements coupling multiple orthogonal unentangled subspaces because the only nonzero 2-RDM elements that couple the ranges $q_n$ and $q_n^\prime$ without implying entangling excitations are equal to the subtrahends of Eqs.\ (\ref{eq:cumulant}) and (\ref{eq:tcumulant}). 

These properties can be applied to Eqs.\ (\ref{eq:Fmat1_cg})--(\ref{eq:Fvecak_cg}) to expose the ranges of explicit ERIs required to calculate gradient and Hessian elements involving given sets of orbital rotations:
\begin{eqnarray}
    \Fmat{1}^{q_1}_{r_1} &=& h^{q_1}_{r_2} D^{r_1}_{r_2} + \{v_\sigma^{(jk)}\}^{q_1}_{r_2} \{\gamma_\sigma\}^{r_1}_{r_2} + g^{q_1r_3}_{r_2r_4} \lambda^{r_1r_3}_{r_2r_4}, \label{eq:Fmat1_c}
    \\
    \Fmat{2}^{q_1s_1}_{r_1t_1} &=& h^{q_1}_{s_1} D^{r_1}_{t_1} + \{v_\sigma^{(jk)}\}^{q_1}_{s_1} \{\gamma_\sigma\}^{r_1}_{t_1} + g^{q_1s_1}_{r_2t_2} \lambda^{r_1t_1}_{r_2t_2} + g^{q_1r_2}_{s_1t_2} \lambda^{r_1r_2}_{t_1t_2} + g^{q_1r_2}_{t_2s_1} \lambda^{r_1r_2}_{t_2t_1}
    \nonumber \\ && + 2g^{q_1s_1}_{r_2t_2}D^{r_1}_{r_2}D^{t_1}_{t_2} - \left(g^{q_1s_1}_{t_2r_2} + g^{q_1t_2}_{s_1r_2}\right)\{\gamma_\sigma\}^{r_1}_{r_2}\{\gamma_\sigma\}^{t_1}_{t_2}, \label{eq:Fmat2_c}
    \\
    \{\Fvec{2}_{\vec{a}_{K1}}\}^{q_1}_{r_1} &=& h^{q_1}_{r_2}\OBMak{D}^{r_1}_{r_2}
    + \{v_\sigma^{(jk)}\}^{q_1}_{r_2}\OBMak{\gamma_\sigma}^{r_1}_{r_2}
    + \OBMak{v_\sigma}^{q_1}_{r_2}\{\gamma_\sigma\}^{r_1}_{r_2}
    + g^{q_1r_3}_{r_2r_4}\OBMak{\lambda}^{r_1r_3}_{r_2r_4}, \label{eq:Fvecak_c}
\end{eqnarray}
where $v_\sigma^{(jk)}$ is defined in the main text and can be evaluated in the AO basis without explicitly transforming ERIs. On the right-hand sides of Eqs.\ (\ref{eq:Fmat1_c})--(\ref{eq:Fvecak_c}), we have specified internal orbital indices using the block-diagonal properties of 1-RDMs and cumulants as discussed above. Comparing Eqs.\ (\ref{eq:Emat1}) and (\ref{eq:Evec2c}) to Eqs.\ (\ref{eq:Fmat1_c}) and (\ref{eq:Fvecak_c}), we see that the evaluation of the derivatives which are first-order in orbital rotation between $\ket{q_n}$ and $\ket{r_n}$ requires, at most, ERIs of the index pattern $g^{q_1r_2}_{r_1r_3}$ and $g^{r_1q_2}_{q_1q_3}$. Comparing Eq.\ (\ref{eq:Emat2}) and (\ref{eq:Fmat2_c}) we see that, for the orbital-orbital Hessian, we require explicit ERIs with the index pattern of the Hessian elements themselves but in all non-equivalent permutations: $\Emat{2}^{p_1r_1}_{q_1s_1}\to g^{p_1r_1}_{q_1s_1},g^{p_1r_1}_{s_1q_1},g^{p_1q_1}_{r_1s_1}$. Even though internal orbital indices appear on the right-hand side of Eq.\ (\ref{eq:Fmat2_c}), they must span the same space as a corresponding external orbital index \emph{if that space is unentangled}.

The above conclusions do not rely on the form of the LAS wave function except inasmuch as $\vec{a}_{K1}$ is defined in Eqs.\ (\ref{eq:tcumulant}) and (\ref{eq:Fvecak_c}); they are otherwise valid for any wave function for which unentangled orbital subspaces can be delineated. The main text uses additional properties of the LAS wave function (specifically, the fact that the cumulant and transition cumulant have nonzero elements only when all four indices span one single active subspace) to further reduce the number of explicitly-evaluated ERIs.

\subsection{Step vector and Hessian-vector products for augmenting the impurity orbitals\label{sec:hessorbs}}

As explained in Sec.\ 2.3.2 of the main text, when constructing the impurity subspace surrounding the $K$th set of active orbitals, the orbital-rotation Hessian is used to augment a set of impurity orbitals by constructing an approximate step vector within the current impurity subspace and evaluating the Hessian-vector product coupling the current impurity to its orthogonal complement. SVD of the Hessian-vector product generates additional impurity orbitals. The approximate step vector is
\begin{eqnarray}
    \vecX^{\tilde{i}_{K1}}_{\tilde{a}_{K1}} &=& \frac{-\Emat{1}^{\tilde{i}_{K1}}_{\tilde{a}_{K1}}}{\Emat{2}^{\tilde{i}_{K1}\tilde{i}_{K1}}_{\tilde{a}_{K1}\tilde{a}_{K1}}}, \label{eq:stepvec}
\end{eqnarray}
where $\ket{\tilde{i}_{Kn}}$ and $\ket{\tilde{a}_{Kn}}$ are linear combinations of impurity orbitals and active orbitals respectively, and are derived from SVD of the orbital rotation gradient:
\begin{eqnarray}
    \Emat{1}^{i_{K1}}_{a_{K1}} &=& \sum_{p_1}^{M_{A_K}} u^{i_{K1}}_{p_1}\sigma_{p_1}\{v^{a_{K1}}_{p_1}\}^*,
    \\
    \{\ket{\tilde{i}_{Kn}}\} &=& \{\ket{i_{K2}}u^{i_{K2}}_{p_1}\},
    \\
    \{\ket{\tilde{a}_{Kn}}\} &=& \{\ket{a_{K2}}v^{a_{K2}}_{p_1}\},
\end{eqnarray}
The implicated second derivatives and Hessian-vector product are
\begin{eqnarray}
    \Emat{2}^{i_{K1}i_{K1}}_{a_{K1}a_{K1}} &=& \Fmat{2}^{i_{K1}i_{K1}}_{a_{K1}a_{K1}} - \Fmat{2}^{i_{K1}a_{K1}}_{a_{K1}i_{K1}} + \Fmat{2}^{a_{K1}a_{K1}}_{i_{K1}i_{K1}}
    - \Fmat{2}^{a_{K1}i_{K1}}_{i_{K1}a_{K1}} - \Fmat{1}^{a_{K1}}_{a_{K1}} - \Fmat{1}^{i_{K1}}_{i_{K1}}, \label{eq:E2iiaa}
    \\
    \EmatX{2}^{u_1}_{i_{K1}} &=& \FmatX{1}^{u_1}_{i_{K1}} - \FmatX{1}^{i_{K1}}_{u_1},
\end{eqnarray}
where, in the ``full Hessian'' protocol,
\begin{eqnarray}
    \Fmat{2}^{i_{K1}i_{K3}}_{i_{K2}i_{K4}} &=& \left(h^{i_{K1}}_{i_{K3}} + \{v_\sigma^{(jk)}\}^{i_{K1}}_{i_{K3}}\right)\{\gamma_\sigma\}^{i_{K2}}_{i_{K4}}
    + g^{i_{K1}i_{K5}}_{i_{K3}i_{K6}}\lambda^{i_{K2}i_{K5}}_{i_{K4}i_{K6}}
    + g^{i_{K1}i_{K3}}_{i_{K5}i_{K6}}\left(\lambda^{i_{K2}i_{K4}}_{i_{K5}i_{K6}} + \lambda^{i_{K2}i_{K6}}_{i_{K5}i_{K4}}\right), \label{eq:Fmat2_imp}
    \\
    \Fmat{1}^{i_{K1}}_{i_{K2}} &=& \left(h^{i_{K1}}_{i_{K3}} + \{v_\sigma^{(jk)}\}^{i_{K1}}_{i_{K3}}\right) \{\gamma_\sigma\}^{i_{K3}}_{i_{K2}}
    + g^{i_{K1}i_{K4}}_{i_{K3}i_{K5}}\lambda^{i_{K2}i_{K4}}_{i_{K3}i_{K5}}, \label{eq:Fmat1_imp}
    \\
    \FmatX{1}^{u_1}_{i_{K1}} &=& \left(h^{u_1}_{i_{K2}} + \{v_\sigma^{(jk)}\}^{u_1}_{i_{K2}}\right) \{\tilde{\gamma}_\sigma\}^{i_{K2}}_{i_{K1}}
    + \{\tilde{v}_\sigma^{(jk)}\}^{u_1}_{i_{K2}} \{\gamma_\sigma\}^{i_{K2}}_{i_{K1}}
    + g^{u_1i_{K3}}_{i_{K2}i_{K4}}\tilde{\lambda}^{i_{K1}i_{K3}}_{i_{K2}i_{K4}}, \label{eq:FmatX2_ui}
    \\
    \FmatX{1}^{i_{K1}}_{u_1} &=& \left(h^{i_{K1}}_{u_2} + \{v_\sigma^{(jk)}\}^{i_{K1}}_{u_2}\right) \{\tilde{\gamma}_\sigma\}^{u_2}_{u_1}
    + \{\tilde{v}_\sigma^{(jk)}\}^{i_{K1}}_{u_2} \{\gamma_\sigma\}^{u_2}_{u_1},
    \label{eq:FmatX2_iu}
\end{eqnarray}
where we have used the fact that both $\ket{\tilde{i}_{Kn}}$ and $\ket{\tilde{a}_{Kn}}$ are contained within $\{\ket{i_{Kn}}\}$. In the ``approximate Hessian'' protocol, Eqs.\ (\ref{eq:Fmat2_imp})--(\ref{eq:FmatX2_iu}) are replaced with
\begin{eqnarray}
    \Fmat{2}^{i_{K1}i_{K3}}_{i_{K2}i_{K4}} &=& \left(h^{i_{K1}}_{i_{K3}} + \{v_\sigma^{(jk)}\}^{i_{K1}}_{i_{K3}}\right)\{\gamma_\sigma\}^{i_{K2}}_{i_{K4}},
    \label{eq:Fmat2_imp_app}
    \\
    \Fmat{1}^{i_{K1}}_{i_{K2}} &=& \left(h^{i_{K1}}_{i_{K3}} + \{v_\sigma^{(jk)}\}^{i_{K1}}_{i_{K3}}\right) \{\gamma_\sigma\}^{i_{K3}}_{i_{K2}},
    \label{eq:Fmat1_imp_app}
    \\
    \FmatX{1}^{u_1}_{i_{K1}} &=& \left(h^{u_1}_{i_{K2}} + \{v_\sigma^{(jk)}\}^{u_1}_{i_{K2}}\right) \{\tilde{\gamma}_\sigma\}^{i_{K2}}_{i_{K1}}
    + \{\tilde{v}_\sigma^{(j)}\}^{u_1}_{i_{K2}} \{\gamma_\sigma\}^{i_{K2}}_{i_{K1}}, \label{eq:FmatX2_ui_app}
    \\
    \FmatX{1}^{i_{K1}}_{u_1} &=& \left(h^{i_{K1}}_{u_2} + \{v_\sigma^{(jk)}\}^{i_{K1}}_{u_2}\right) \{\tilde{\gamma}_\sigma\}^{u_2}_{u_1}
    + \{\tilde{v}_\sigma^{(j)}\}^{i_{K1}}_{u_2} \{\gamma_\sigma\}^{u_2}_{u_1},
    \label{eq:FmatX2_iu_app}
\end{eqnarray}
using
\begin{eqnarray}
    \{\tilde{v}^{(j)}\}^{\mu_1}_{\mu_2} &=& g^{\mu_1\mu_3}_{\mu_2\mu_4}\tilde{D}^{\mu_3}_{\mu_4},
\end{eqnarray}
which is just the (fast) Coulomb part of an effective potential evaluated in the AO basis, omitting the (slow) exchange part.

\section{Cartesian coordinates} 

\begin{table}[H]
\caption{System a) (Fe[NCH]$_6^{2+}$) molecular geometry in $\AA$, taken from Ref.\ \citenum{Pandharkar2019}}
\begin{tabular}{|l|r|r|r|}
\hline & \multicolumn{1}{c|}{x} & \multicolumn{1}{c|}{y} & \multicolumn{1}{c|}{z} \\ \hline
Fe & 0.000000 & 0.000000 & 0.000000 \\ \hline 
N & 0.000000 & 0.000000 & 1.949733 \\ \hline 
N & 0.000000 & 1.949733 & -0.000000 \\ \hline 
N & -1.949733 & 0.000000 & -0.000000 \\ \hline 
N & 0.000000 & -1.949733 & 0.000000 \\ \hline 
N & 0.000000 & -0.000000 & -1.949733 \\ \hline 
N & 1.949733 & -0.000000 & 0.000000 \\ \hline 
C & 0.000000 & 0.000000 & 3.090409 \\ \hline 
H & 0.000000 & 0.000000 & 4.168181 \\ \hline 
C & -3.090409 & 0.000000 & -0.000000 \\ \hline 
H & -4.168181 & 0.000000 & -0.000000 \\ \hline 
C & 0.000000 & -3.090409 & 0.000000 \\ \hline 
H & 0.000000 & -4.168181 & 0.000000 \\ \hline 
C & 0.000000 & 3.090409 & -0.000000 \\ \hline 
H & 0.000000 & 4.168181 & -0.000000 \\ \hline 
C & 3.090409 & -0.000000 & 0.000000 \\ \hline 
H & 4.168181 & -0.000000 & 0.000000 \\ \hline 
C & 0.000000 & -0.000000 & -3.090409 \\ \hline 
H & 0.000000 & -0.000000 & -4.168181 \\ \hline 
\end{tabular}
\label{fench_geom}
\end{table}

\begin{table}[H]
\caption{System b) (bisdiazene) B3LYP/6-31g(d,p) equilibrium molecular geometry in $\AA$, taken from Ref.\ \citenum{Hermes2019}}
\begin{tabular}{|l|r|r|r|}
\multicolumn{4}{l}{$E=-298.713635\ E_\textrm{h}$} \\ \hline
 & \multicolumn{1}{c|}{x} & \multicolumn{1}{c|}{y} & \multicolumn{1}{c|}{z} \\ \hline
H & -0.145525 & 2.534624 & 1.196098 \\ \hline
N & 0.586282 & 2.685058 & 0.454251 \\ \hline
N & 0.586282 & 1.768962 & -0.376701 \\ \hline
C & -0.376894 & 0.666619 & -0.222182 \\ \hline
H & -0.986448 & 0.788749 & 0.689250 \\ \hline
H & -1.039496 & 0.688384 & -1.095107 \\ \hline
C & 0.376894 & -0.666619 & -0.222182 \\ \hline
H & 0.986448 & -0.788749 & 0.689250 \\ \hline
H & 1.039496 & -0.688384 & -1.095107 \\ \hline
N & -0.586282 & -1.768962 & -0.376701 \\ \hline
N & -0.586282 & -2.685058 & 0.454251 \\ \hline
H & 0.145525 & -2.534624 & 1.196098 \\ \hline
\end{tabular}
\label{c2h6n4_geom}
\end{table}

\begin{table}[H]
\caption{System c) (high-spin all-\emph{trans} polyacetylene) molecular geometry in $\AA$ for the case of $n=1$, adapted from Ref.\ \citenum{Hirata2004}.}
\begin{tabular}{|l|r|r|r|}
\hline  & \multicolumn{1}{c|}{x} & \multicolumn{1}{c|}{y} & \multicolumn{1}{c|}{z} \\ \hline
H & -0.517230236100 & -0.960600792700 & 0.000000000000 \\ \hline 
C & 0.000000000000 & 0.000000000000 & 0.000000000000 \\ \hline 
H & -0.517230236100 & 0.960600792700 & 0.000000000000 \\ \hline 
C & 1.369000000000 & 0.000000000000 & 0.000000000000 \\ \hline 
H & 1.886230236100 & -0.960600792700 & 0.000000000000 \\ \hline 
C & 2.176695293900 & 1.175203945000 & 0.000000000000 \\ \hline 
H & 1.659465057800 & 2.135804737700 & 0.000000000000 \\ \hline 
C & 3.545695293900 & 1.175203945000 & 0.000000000000 \\ \hline 
H & 4.062925529900 & 0.214603152300 & 0.000000000000 \\ \hline 
C & 4.353390587700 & 2.350407889900 & 0.000000000000 \\ \hline 
H & 3.836160351600 & 3.311008682600 & 0.000000000000 \\ \hline 
C & 5.722390587700 & 2.350407889900 & 0.000000000000 \\ \hline 
H & 6.239620823800 & 1.389807097300 & 0.000000000000 \\ \hline 
H & 6.239620823800 & 3.311008682600 & 0.000000000000 \\ \hline 
\end{tabular}
\label{polyacet_geom}
\end{table}

\section{Raw data}

\begin{table}[H]
\caption{Convergence over successive macrocycles (i.e., loops through all four decision nodes of the algorithms depicted in Figs.\ 2 and 3 of the main text) of the singlet electronic energy of [Fe(NCH)$_6$]$^{2+}$ with the LASSCF(6,5)/ANO-RCC-VTZP and vLASSCF(6,5)/ANO-RCC-VTZP methods using various schemes for constructing impurity orbitals in the latter, compared to the        CASSCF(6,5)/ANO-RCC-VTZP singlet energy. Calculations initialized from the orbitals provided by a low-spin RHF wave function. \textsuperscript{\emph{a}}The energy reaches fixed point to within $10^{-6}\ E_{\mathrm{h}}$, but the gradient norm still $>10^{-4}$, therefore the calculation is not converged.}
\begin{tabular}{|r|c|c|c|c|}
\multicolumn{5}{c}{$E_{\mathrm{CAS}} = -1828.6865336\ E_{\mathrm{h}}$} \\ \hline
\multirow{2}{*}{SCF cycle} & \multirow{2}{*}{$E_{\mathrm{LAS}}$} & \multicolumn{3}{|c|}{$E_{\mathrm{vLAS}}$} \\ \cline{3-5}
  & & no Hessian & approximate Hessian & full Hessian \\ \hline
1 & -1828.683021 & -1828.683021 & -1828.683021 & -1828.683021 \\ \hline
2 & -1828.685940 & -1828.686353 & -1828.686451 & -1828.686441 \\ \hline
3 & -1828.685940 & -1828.686457 & -1828.686512 & -1828.686508 \\ \hline
4 & -1828.685940 & -1828.686483 & -1828.686526 & -1828.686524 \\ \hline
5 & -1828.685940 & -1828.686497 & -1828.686531 & -1828.686530 \\ \hline
6 & ? & -1828.686507 & -1828.686533 & -1828.686533 \\ \hline
7 & ? & -1828.686513 & -1828.686533 & -1828.686533 \\ \hline
8 & ? & -1828.686518 & -1828.686533 & -1828.686533 \\ \hline
9 & ? & -1828.686521 & -1828.686533 & -1828.686533 \\ \hline
10 & ? & -1828.686524 & -1828.686533 & ? \\ \hline
11 & ? & -1828.686526 & -1828.686533 & ? \\ \hline
12 & ? & -1828.686528 & ? & ? \\ \hline
13 & ? & -1828.686529 & ? & ? \\ \hline
14 & ? & -1828.686530 & ? & ? \\ \hline
15 & ? & -1828.686531 & ? & ? \\ \hline
16 & ? & -1828.686531 & ? & ? \\ \hline
17 & ? & -1828.686532 & ? & ? \\ \hline
18 & ? & -1828.686532 & ? & ? \\ \hline
19 & ? & -1828.686532 & ? & ? \\ \hline
20 & ? & -1828.686533 & ? & ? \\ \hline
21 & ? & -1828.686533\textsuperscript{\emph{a}} & ? & ? \\ \hline
\end{tabular}
\label{fig4_1}
\end{table}

\begin{table}[H]
\caption{Convergence over successive macrocycles of the singlet electronic energy of [Fe(NCH)$_6$]$^{2+}$ with the LASSCF(6,5)/ANO-RCC-VTZP and vLASSCF(6,5)/ANO-RCC-VTZP methods using various schemes for constructing impurity orbitals in the latter, compared to the        CASSCF(6,5)/ANO-RCC-VTZP singlet energy. Calculations initialized from the orbitals provided by a high-spin (quintet, $s=2$) ROHF wave function. \textsuperscript{\emph{a}}LASSCF and vLASSCF with no Hessian are not converged after 30 cycles.}
\begin{tabular}{|r|c|c|c|c|}
\multicolumn{5}{c}{$E_{\mathrm{CAS}} = -1828.6865336\ E_{\mathrm{h}}$} \\ \hline
\multirow{2}{*}{SCF cycle} & \multirow{2}{*}{$E_{\mathrm{LAS}}$} & \multicolumn{3}{|c|}{$E_{\mathrm{vLAS}}$} \\ \cline{3-5}
  & & no Hessian & approximate Hessian & full Hessian \\ \hline
1 & -1828.676217 & -1828.676218 & -1828.676218 & -1828.676218 \\ \hline
2 & -1828.682773 & -1828.684798 & -1828.685699 & -1828.685493 \\ \hline
3 & -1828.682780 & -1828.685540 & -1828.686296 & -1828.686218 \\ \hline
4 & -1828.682782 & -1828.685811 & -1828.686453 & -1828.686438 \\ \hline
5 & -1828.682782 & -1828.685981 & -1828.686504 & -1828.686509 \\ \hline
6 & -1828.682782 & -1828.686106 & -1828.686525 & -1828.686527 \\ \hline
7 & -1828.682783 & -1828.686199 & -1828.686531 & -1828.686531 \\ \hline
8 & -1828.682783 & -1828.686271 & -1828.686533 & -1828.686533 \\ \hline
9 & -1828.682783 & -1828.686326 & -1828.686533 & -1828.686533 \\ \hline
10 & -1828.682783 & -1828.686369 & -1828.686533 & -1828.686533 \\ \hline
11 & -1828.682783 & -1828.686403 & -1828.686534 & -1828.686534 \\ \hline
12 & -1828.682782 & -1828.686430 & -1828.686534 & -1828.686534 \\ \hline
13 & -1828.682782 & -1828.686451 & -1828.686534 & -1828.686534 \\ \hline
14 & -1828.682782 & -1828.686468 & -1828.686534 & -1828.686534 \\ \hline
15 & -1828.682782 & -1828.686481 & ? & -1828.686534 \\ \hline
16 & -1828.682782 & -1828.686492 & ? & -1828.686534 \\ \hline
17 & -1828.682782 & -1828.686500 & ? & ? \\ \hline
18 & -1828.682782 & -1828.686507 & ? & ? \\ \hline
19 & -1828.682781 & -1828.686512 & ? & ? \\ \hline
20 & -1828.682781 & -1828.686516 & ? & ? \\ \hline
21 & -1828.682780 & -1828.686520 & ? & ? \\ \hline
22 & -1828.682779 & -1828.686523 & ? & ? \\ \hline
23 & -1828.682778 & -1828.686525 & ? & ? \\ \hline
24 & -1828.682777 & -1828.686527 & ? & ? \\ \hline
25 & -1828.682775 & -1828.686528 & ? & ? \\ \hline
26 & -1828.682772 & -1828.686529 & ? & ? \\ \hline
27 & -1828.682769 & -1828.686530 & ? & ? \\ \hline
28 & -1828.682764 & -1828.686531 & ? & ? \\ \hline
29 & -1828.682783 & -1828.686531 & ? & ? \\ \hline
30\textsuperscript{\emph{a}} & -1828.682783 & -1828.686532 & ? & ? \\ \hline
\end{tabular}
\label{fig4_2}
\end{table}

\begin{table}[H]
\caption{Potential energy curves of system b) (bisdiazene) under simultaneous double-bond dissociation calculated with CASSCF(8,8), LASSCF(8,8), and vLASSCF(8,8) in the 6-31g basis for $R_{\mathrm{NN}}\le 5.04\ \AA$. \textsuperscript{\emph{a}}From Ref.\ \citenum{Hermes2019}. \textsuperscript{\emph{b}}B3LYP/6-31g equilibrium geometry, i.e., $R_{\mathrm{NN}}=1.236816\ \AA$.
\textsuperscript{\emph{c}}Omitted from the plot in Ref.\ \citenum{Hermes2019} as suspected numerical error as indicated in that paper's SI (but not omitted from Fig.\ 6 of this work's main text).}
\begin{tabular}{|r|c|c|c|}
\hline
$R_{\mathrm{NN}} / \AA$ & CASSCF & LASSCF\textsuperscript{\emph{a}} & vLASSCF \\ \hline
0.94 & -296.255073 & -296.248733 & -296.255058 \\ \hline
1.04 & -296.644073 & -296.643922 & -296.644051 \\ \hline
1.14 & -296.820276 & -296.820078 & -296.820244 \\ \hline
1.24\textsuperscript{\emph{b}} & -296.879579 & -296.879306 & -296.879530 \\ \hline
1.34 & -296.877791 & -296.877509 & -296.877718 \\ \hline
1.44 & -296.847102 & -296.846649 & -296.846996 \\ \hline
1.54 & -296.805552 & -296.804856 & -296.805400 \\ \hline
1.64 & -296.762823 & -296.761870 & -296.762608 \\ \hline
1.74 & -296.723935 & -296.722753 & -296.723639 \\ \hline
1.84 & -296.691474 & -296.690120 & -296.691080 \\ \hline
1.94 & -296.666650 & -296.665195 & -296.666146 \\ \hline
2.04 & -296.649552 & -296.648091 & -296.648944 \\ \hline
2.14 & -296.639182 & -296.637617 & -296.638498 \\ \hline
2.24 & -296.633801 & -296.632159 & -296.633074 \\ \hline
2.34 & -296.631582 & -296.629891 & -296.630840 \\ \hline
2.44 & -296.631109 & -296.629389 & -296.630366 \\ \hline
2.54 & -296.631486 & -296.629751 & -296.630748 \\ \hline
2.64 & -296.632218 & -296.630483 & -296.631483 \\ \hline
2.74 & -296.633057 & -296.631368 & -296.632320 \\ \hline
2.84 & -296.633890 & -296.632178 & -296.633143 \\ \hline
2.94 & -296.634671 & -296.632928 & -296.633902 \\ \hline
3.04 & -296.635387 & -296.633604 & -296.634580 \\ \hline
3.14 & -296.636032 & -296.632168 & -296.635171 \\ \hline
3.24 & -296.636600 & -296.632712 & -296.635677 \\ \hline
3.34 & -296.637090 & -296.632491 & -296.636102 \\ \hline
3.44 & -296.637505 & -296.632893 & -296.636455 \\ \hline
3.54 & -296.637855 & -296.629952\textsuperscript{\emph{c}} & -296.636746 \\ \hline
3.64 & -296.638149 & -296.634377 & -296.636987 \\ \hline
3.74 & -296.638395 & -296.627817\textsuperscript{\emph{c}} & -296.637188 \\ \hline
3.84 & -296.638604 & -296.634648 & -296.637361 \\ \hline
3.94 & -296.638784 & -296.634771 & -296.637513 \\ \hline
4.04 & -296.638942 & -296.634879 & -296.637651 \\ \hline
4.14 & -296.639085 & -296.635080 & -296.637779 \\ \hline
4.24 & -296.639217 & -296.635353 & -296.637900 \\ \hline
4.44 & -296.639454 & ? & -296.638126 \\ \hline
4.64 & -296.639664 & ? & -296.638332 \\ \hline
4.84 & -296.639849 & ? & -296.638516 \\ \hline
5.04 & -296.640011 & ? & -296.638679 \\ \hline
\end{tabular}
\label{fig5_1}
\end{table}

\begin{table}[H]
\caption{Potential energy curves of system b) (bisdiazene) under simultaneous double-bond dissociation calculated with CASSCF(8,8) and vLASSCF(8,8) in the 6-31g basis for geometries $R_{\mathrm{NN}}>5.04\ \AA$. \textsuperscript{\emph{a}}Dissociation limit: CASSCF and vLASSCF energies converged to $<10^{-6} E_{\mathrm{h}}$.}
\begin{tabular}{|r|c|c|}
\hline
$R_{\mathrm{NN}} / \AA$ & CASSCF & vLASSCF \\ \hline
5.24 & -296.640151 & -296.638820 \\ \hline
5.44 & -296.640273 & -296.638943 \\ \hline
5.64 & -296.640378 & -296.639049 \\ \hline
5.84 & -296.640470 & -296.639142 \\ \hline
6.04 & -296.640549 & -296.639222 \\ \hline
6.24 & -296.640619 & -296.639292 \\ \hline
6.44 & -296.640680 & -296.639354 \\ \hline
6.64 & -296.640734 & -296.639408 \\ \hline
6.84 & -296.640781 & -296.639456 \\ \hline
7.04 & -296.640823 & -296.639499 \\ \hline
7.24 & -296.640861 & -296.639536 \\ \hline
7.44 & -296.640894 & -296.639570 \\ \hline
7.64 & -296.640924 & -296.639599 \\ \hline
7.84 & -296.640950 & -296.639626 \\ \hline
8.04 & -296.640974 & -296.639650 \\ \hline
8.24 & -296.640995 & -296.639672 \\ \hline
8.44 & -296.641015 & -296.639691 \\ \hline
8.64 & -296.641032 & -296.639709 \\ \hline
8.84 & -296.641048 & -296.639725 \\ \hline
9.04 & -296.641062 & -296.639739 \\ \hline
9.24 & -296.641075 & -296.639752 \\ \hline
9.44 & -296.641087 & -296.639764 \\ \hline
9.64 & -296.641098 & -296.639775 \\ \hline
9.84 & -296.641108 & -296.639785 \\ \hline
10.04 & -296.641117 & -296.639794 \\ \hline
10.24 & -296.641125 & -296.639802 \\ \hline
10.44 & -296.641132 & -296.639810 \\ \hline
10.64 & -296.641139 & -296.639817 \\ \hline
10.84 & -296.641146 & -296.639823 \\ \hline
11.04 & -296.641152 & -296.639829 \\ \hline
11.24 & -296.641157 & -296.639834 \\ \hline
16.24 & -296.641217 & -296.639895 \\ \hline
21.24 & -296.641231 & -296.639909 \\ \hline
26.24 & -296.641235 & -296.639913 \\ \hline
31.24 & -296.641237 & -296.639915 \\ \hline
41.24 & -296.641238 & -296.639916 \\ \hline
56.24\textsuperscript{\emph{a}} & -296.641239 & -296.639917 \\ \hline
101.24 & -296.641239 & -296.639917 \\ \hline
\end{tabular}
\label{fig5_2}
\end{table}

\begin{table}[H]
\caption{Wall time in seconds required to complete CASSCF, LASSCF, and vLASSCF calculations of high-spin states of $(n+2)$-polyene chains with $1\le n<22$ in the 6-31g basis compared to $M_{\mathrm{tot}}$ (the number of atomic orbitals) using one thread and 62 GB of memory. \textsuperscript{\emph{a}}Program crashed for unknown reason.}
\begin{tabular}{|r|c|c|c|c|}
\hline
\multirow{2}{*}{$M_{\mathrm{tot}}$} & \multirow{2}{*}{CASSCF} & \multirow{2}{*}{LASSCF} & \multicolumn{2}{|c|}{vLASSCF} \\ \cline{4-5}
& & & approximate Hessian & full Hessian \\ \hline
70 & 3 & 19 & 21 & 26 \\ \hline
92 & 7 & 24 & 37 & 41 \\ \hline
114 & 15 & 40 & 45 & 61 \\ \hline
136 & 27 & 44 & 62 & 84 \\ \hline
158 & 51 & 58 & 85 & 116 \\ \hline
180 & 77 & 77 & 119 & 161 \\ \hline
202 & 121 & 107 & 165 & 240 \\ \hline
224 & 181 & 135 & 197 & 322 \\ \hline
246 & 243 & 175 & 259 & 470 \\ \hline
268 & 350 & 221 & 332 & 587 \\ \hline
290 & 510 & 295 & 438 & 808 \\ \hline
312 & 653 & 363 & 545 & 967 \\ \hline
334 & 864 & 495 & 694 & 1262 \\ \hline
356 & 1168 & 571 & 923 & 1619 \\ \hline
378 & 1488 & 742 & 1139 & ? \\ \hline
400 & 1856 & 908 & 1398 & ? \\ \hline
422 & 2287 & 1088 & 1663 & 2847 \\ \hline
444 & 2971 & 1356 & 1999 & 3181 \\ \hline
466 & 3540 & 1514 & 2298 & 4009 \\ \hline
488 & 4184 & 1751 & 2698 & 4832 \\ \hline
510 & 5100 & 2190 & 3245 & 5650 \\ \hline
\end{tabular}
\label{fig6_1}
\end{table}

\begin{table}[H]
\caption{Average wall time in seconds required to calculate the Coulomb and exchange effective potential matrices using the two available standard implementations in {\sc PySCF} in the CASSCF calculations reported in Table \ref{fig6_1}.}
\begin{tabular}{|r|c|c|}
\hline
$M_{\mathrm{tot}}$ & Fast implementation & Slow implementation \\ \hline
70 & 0.04 & 0.08 \\ \hline
92 & 0.11 & 0.22 \\ \hline
114 & 0.24 & 0.48 \\ \hline
136 & 0.46 & 0.89 \\ \hline
158 & 0.81 & 1.64 \\ \hline
180 & 1.26 & 2.48 \\ \hline
202 & 2.01 & 3.99 \\ \hline
224 & 2.91 & 5.82 \\ \hline
246 & 4.09 & 8.12 \\ \hline
268 & 5.60 & 11.16 \\ \hline
290 & 7.52 & 15.00 \\ \hline
312 & 9.98 & 19.91 \\ \hline
334 & 12.93 & 25.70 \\ \hline
356 & 16.89 & 33.67 \\ \hline
378 & 21.35 & 42.59 \\ \hline
400 & 25.67 & 52.01 \\ \hline
422 & 31.23 & 63.69 \\ \hline
444 & 38.98 & 78.19 \\ \hline
466 & 46.92 & 95.07 \\ \hline
488 & 54.90 & 110.91 \\ \hline
510 & 66.14 & 133.14 \\ \hline
\end{tabular}
\label{fig6_2}
\end{table}

\begin{table}[H]
\caption{Total electronic energies from the calculations reported in Table \ref{fig6_1}. \textsuperscript{\emph{a}}Program crashed for unknown reason.}
\begin{tabular}{|r|c|c|c|c|}
\hline
\multirow{2}{*}{$M_{\mathrm{tot}}$} & \multirow{2}{*}{CASSCF} & \multirow{2}{*}{LASSCF} & \multicolumn{2}{|c|}{vLASSCF} \\ \cline{4-5}
& & & approximate Hessian & full Hessian \\ \hline
70 & -231.275657 & -231.269921 & -231.275657 & -231.275657 \\ \hline
92 & -307.972626 & -307.960879 & -307.972626 & -307.972626 \\ \hline
114 & -384.669592 & -384.653153 & -384.669592 & -384.669592 \\ \hline
136 & -461.366558 & -461.341623 & -461.366558 & -461.366558 \\ \hline
158 & -538.063525 & -538.032551 & -538.063525 & -538.063525 \\ \hline
180 & -614.760491 & -614.723532 & -614.760491 & -614.760491 \\ \hline
202 & -691.457458 & -691.414558 & -691.457458 & -691.457458 \\ \hline
224 & -768.154425 & -768.105604 & -768.154425 & -768.154425 \\ \hline
246 & -844.851392 & -844.796658 & -844.851392 & -844.851392 \\ \hline
268 & -921.548359 & -921.487717 & -921.548359 & -921.548359 \\ \hline
290 & -998.245325 & -998.178778 & -998.245325 & -998.245325 \\ \hline
312 & -1074.942292 & -1074.869840 & -1074.942292 & -1074.942292 \\ \hline
334 & -1151.639259 & -1151.560902 & -1151.639259 & -1151.639259 \\ \hline
356 & -1228.336226 & -1228.251964 & -1228.336226 & -1228.336226 \\ \hline
378 & -1305.033193 & -1304.943027 & -1305.033193 & ?\textsuperscript{\emph{a}} \\ \hline
400 & -1381.730159 & -1381.634089 & -1381.730159 & ?\textsuperscript{\emph{a}} \\ \hline
422 & -1458.427126 & -1458.325152 & -1458.427126 & -1458.427126 \\ \hline
444 & -1535.124093 & -1535.016214 & -1535.124093 & -1535.124093 \\ \hline
466 & -1611.821060 & -1611.707277 & -1611.821060 & -1611.821060 \\ \hline
488 & -1688.518027 & -1688.398339 & -1688.518027 & -1688.518027 \\ \hline
510 & -1765.214994 & -1765.089402 & -1765.214994 & -1765.214994 \\ \hline
\end{tabular}
\label{fig6_3}
\end{table}

\bibliography{draft3.bib}